
Please, find below the TEX file for the text, the TEX file
for the Tables, the TEX file for the first two figures and
the POSTCRIPT file for the third figure.

*********************************************************

\magnification=1200
\baselineskip=19truept

\leftskip=27truepc
\noindent UG--FT--38/94\hfil\break
\noindent  OUTP--94--6P\hfil\break
\noindent May 1994\hfil\break

\rightskip=3truepc
\leftskip=3truepc

\vskip 0.3truecm
\centerline{\bf PATTERNS OF QUARK MASS MATRICES}
\centerline{\bf IN A CLASS OF CALABI-YAU MODELS }
\vskip 0.5truecm
\centerline{F. del Aguila$^{(a)}$, M. Masip$^{(a)}$,
L. da Mota$^{(b)}$}
\vskip .5truecm
\centerline{\it $^{(a)}$Departamento de F\'\i sica
Te\'orica y del Cosmos}
\centerline{\it Universidad de Granada}
\centerline{\it 18071 Granada, Spain}
\vskip .3truecm
\centerline{\it $^{(b)}$Department of Theoretical Physics}
\centerline{\it University of Oxford}
\centerline{\it 1 Keble Road, Oxford OX1 3NP, UK}
\vskip 0.8truecm

\centerline{ABSTRACT}
\vskip 0.1truecm

We study a class of superstring
models compactified in the 3-generation
Calabi-Yau manifold of Tian and Yau. Our analysis includes
the complete $E_6$-singlet sector,
which has been recently evaluated
using techniques of spectral and exact sequences.
We use the discrete symmetries of
the models to find flat
directions of symmetry breaking that leave unbroken a
low energy matter parity and make all leptoquarks
heavy while preserving light Higgs fields. Then we classify
the patterns of ordinary quark mass matrices and
show that (without invoking effects due to nonrenormalizable
terms) only one
structure can accommodate the observed value of fermion masses
and mixing angles, with preference for a heavy {\it top}
quark ( $m_t\ge 170$ GeV for $V_{13}\le 0.013$ ).
The model, which unifies perturbatively and predicts a realistic
structure of quark mass matrices with texture zeroes, is
one of the many possible string vacua. However,
in contrast with what is often assumed in the search for
realistic unified scenarios, it is highly nonminimal
near the unification scale and the predicted mass matrices
have no simple symmetry properties.

\vfil
\eject

\leftskip=2.pc
\rightskip=1.pc
\baselineskip=22truept

\vskip 0.3truecm
\centerline{\bf I. INTRODUCTION}
\vskip 0.2truecm

{}From the appearance of grand unified theories,
the idea of unification of gauge couplings and the
possibility of understanding the assignment of standard
quantum numbers, the replication of families, and the pattern
of fermion masses and mixing angles have been generic requirements
in the search for a fundamental theory at very large energy
scales [1]. The stability of the
different mass scales involved (the electroweak
and the unification scales) seems to demand a supersymmetric
model.
In addition, the proximity of the unification and the Planck
scales points to the necessity of including gravity. String
theory [2] has the potential of realizing
all these ideas consistently.
Hence, a program for obtaining the standard model from the
string was proposed [3]. However, it is not known how to
derive the physical vacuum from the string.
Alternatively, one can scan compactified string models
requiring consistency with the observed phenomenology.
Possible candidates are the models resulting from the
ten dimensional $E_8\times E_8$
heterotic superstring compactified
on the 3-generation Calabi-Yau manifold of Tian
and Yau [4].
In this paper we analyze the class of
Tian-Yau models with a gauge symmetry
$$G=SU(3)_C\times SU(3)_L\times SU(3)_R\subset E_6\eqno(1)$$
and a group of discrete symmetries of order 72 [5] (see
Section 1.1).
We use the available techniques to study this class
of models quantitatively in some detail and, as a result,
we single out only one realistic case. Our approach goes
{\it up-down} from the compactification to the electroweak
scale.
It is remarkable
that, in contrast with the phenomenological approach going
{\it down-up},
the model found is
highly nonminimal near the unification scale. Hence, this
is an example of a consistent model derived from the string
which does not
share the simplicity and predictivity
of the minimal supersymmetric grand unified scenarios [6,7].
The perturbative unification takes place with a nonminimal
matter content and after several
intermediate scales of symmetry breaking. Moreover,
the fermion mass matrices, although with a
perturbative structure and texture zeroes, are not
simple (symmetric).

The phenomenological aspects
of this class of models have been
discussed previously, and
the possibility of obtaining an almost
 minimal supersymmetric standard model
at low energy and a realistic
pattern of fermion masses and mixings emphasized
[5,8]. This analysis
made use of unknown soft supersymmetry breaking terms,
nonrenormalizable terms, and $E_6$-singlet fields. Recently the
$E_6$-singlet sector has been  calculated using cohomology
techniques of spectral sequences [9].
Here we reanalyse the low energy predictions
 of these models when
the complete singlet and nonsinglet spectra
and their renormalizable couplings at the compactification
scale are included.
We will assume that supersymmetry breaking
[10] occurs
and generates the required soft breaking terms. Nonrenormalizable
terms [11] have not been worked out. We
assume that they are largely suppressed and, in consequence,
do not spoil the flatness conditions and
introduce at most order 1 TeV mass contributions for the
different fields.
We search for realistic models
requiring that the intermediate
scales of symmetry breaking (see next section)
\item {\it (i)} preserve a low-energy matter parity,
\item {\it (ii)} give heavy masses to all the leptoquarks,
\item {\it (iii)}  keep the standard Higgs fields light, and
\item {\it (iv)} take place along flat directions of the potential.

\noindent
Then we work out the low-energy spectrum and the ordinary
fermion mass matrix predictions. We come out with
 4 {\it types} of structures, but only one
can accommodate the observed
quark masses and the Cabibbo-Kobayashi-Maskawa matrix
$V_{ij}$.
Ramond, Roberts, and Ross (RRR) [7] have stitched the Yukawa quilt
classifying the possible patterns of symmetric (or hermitian)
quark mass matrices  with texture zeroes. These matrices
incorporate the perturbative structure observed
in the quark sector: a heavy third family and the rest of masses
expressed as an expanssion in powers of $V_{12}$.
As they point out, these matrices  are a natural expectation
in compactified string theory with broken discrete symmetries.
The four {\it types} ({\it A, B, C, D})
of patterns that we find are explicit
realizations of their mechanism. We do not find, however,
any of the RRR textures. Our matrices are not
necessarily symmetric and contain less texture zeroes.

In what follows we present the class of models, the two
$Z_N$ matter parities, $P_{2,3}$, which can be defined in this
class, and illustrate
the type of perturbative structures and texture
zeroes which give nontrivial constraints on fermion masses
and mixing angles. In Section II (III) we analyse the
models with $P_3 (P_2)$ matter parity. We argue that only the
model with mass matrices of {\it type D} is phenomenologically
viable and discuss the flatness conditions for this model is
some detail. Section IV is devoted to conclusions.

\vskip 0.3truecm
\noindent 1.1. THE CLASS OF MODELS
\vskip 0.2truecm

The Tian-Yau manifold under study [5,8,9] is constructed as the
quotient manifold $R/G$, where  $R$ is the space of solutions
in $CP^{3}_x\times CP^{3}_y$ to
$$\eqalign{ f(x)&=\sum\limits_{i=1}^{4} (x^i)^3 = 0\; ;\; \;
g(y)=\sum\limits_{i=1}^{4} (y^i)^3 = 0\; ;\cr
h(x,y)&= x^1y^1 +x^2y^2 +c\; (\; x^3y^3 +x^4y^4\; )  = 0\; ,\cr}
\eqno(2)$$
and $G$ is the freely-acting $Z_3$ discrete group
$$g:(x^1,x^2,x^3,x^4;y^1,y^2,y^3,y^4)\rightarrow
(x^1,\alpha^2x^2,\alpha x^3,\alpha x^4;
y^1,\alpha y^2,\alpha^2y^3,\alpha^2y^4)\ .
\eqno(3)$$
This manifold
defines a variety of unified models of electroweak
and strong interactions with 3 chiral generations.
Our choice of the complex structure (which is given
in Eq.~(2) in terms of
the arbitrary parameter $c$) implies an order 72 group of
discrete symmetries [5,8].
The nonsinglet matter content
at the compactification scale $M_c$ consists of nine
families of leptons
$\lambda$, six of $\overline \lambda$, seven of quarks $q$
and antiquarks $Q$, and four of $\overline q$ and $\overline Q$,
where
$${\bf 27}\longrightarrow \ ({\bf 1},\overline {\bf 3},
{\bf 3})\equiv\lambda\ +\
({\bf 3},{\bf 3},{\bf 1})\equiv q\ +\ (\overline {\bf 3},
{\bf 1}, \overline {\bf 3}) \equiv Q
\eqno(4)$$
and the assignment of standard quantum numbers is
$$\lambda\sim\pmatrix{
h^0&h'^-&e\cr
h^+&h'^0&\nu\cr
e^c&\nu^c&N\cr};\
q\sim\pmatrix{
u&\cr
d&\times\ 3\ colors\cr
d'&\cr};\
Q\sim\pmatrix{
u^c&\cr
d^c&\times\ 3\ colors\cr
d'^c&\cr}.\eqno(5)$$
The transformation properties of these fields
under the group of discrete symmetries
are listed in Table I.

The $E_6$-singlet sector of the Tian-Yau construction
has been recently
calculated by Hubsch and collaborators using
the cohomology techniques of exact and spectral sequences [9].
These methods generalize previous results obtained via
polinomial deformations, providing a parametrization
of all the ${\bf 27}$, $\overline {\bf 27}$, and ${\bf 1}$
fields
and a framework to analyze their couplings.
The models with the complex
structure that we have chosen contain 20 singlets
(this number jumps for special
choices of the manifold, like in the R-symmetric case
considered in Ref.[9]). Their tensor representatives
and transformation properties are listed in Table II
(the fields in Tables I,II correspond to $B$
eigenstates; we can also
choose a basis of $C$ eigenstates, as in Table III).
The masses that these fields
receive through instanton corrections [11]
may be consistently small,
and, at any rate,  their ${\bf 1\ 27\ \overline{27}}$
couplings with the nonsinglet sector make them an essential
part of the model at
 $M_c$. We should remark that among the singlets calculated
in Ref.~[9]
there is {\it not} a subset with transformation properties
identical to those of $\lambda$ and $\overline \lambda$
fields. (In the past it was usually assumed, based on theorems
on the dimension of various cohomology groups on the manifold,
a one to one correspondence
between ${\bf 27}'s$ and some $E_6$ singlets.)

Once the supersymmetry breaking terms are included,
 the effective theory
is specified if we know the dominant terms in
the superpotential. The trilinear terms must be invariant
under the discrete symmetries of the model, and
at first order in $\sigma$ perturbation theory
can be calculated in terms of the
parameter $c$ specifying the complex structure of the
manifold [8,9], up to normalization factors of the fields.
We write in Tables IV and V all
the nonzero trilinears of type $\lambda^3$,
$\lambda q Q$ and $s \lambda \overline \lambda$, $s^3$,
respectively.
These are the relevant terms in the discussion of fermion
masses and flat directions.
Due to our lack of knowledge of the field normalization
factors, the
explicit expression of all the coefficients
is not necessary.
The symmetries of the manifold also restrict the
possible nonrenormalizable terms, which appear nonperturbatively
after integrating out the massive string and Kaluza-Klein modes.
In the present work we will
use large
singlet VEVs to make massive many of the vectorlike
families that appear at $M_c$. This enables us to
consistently {\it neglect} all nonrenormalizable terms.
(The suppression could be justified by their
nonperturbative origin.) The motivation to do this
is phenomenological, since it has been shown that
nonrenormalizable terms tend to give too large
masses to the standard Higgses [12] and to spoil the flatness
conditions [13]. Both problems could be avoided
by the action of an exact discrete symmetry [13,14],
but in these models
unbroken symmetries imply the presence of extra
light fields giving an unacceptable value for the
proton lifetime and/or low energy gauge couplings [15].

The rank-6 models under discussion
require two large intermediate scales of
gauge symmetry breaking defined by VEVs
along $N$ and $\nu^c$ (plus identical
$\overline N$ and $\overline \nu^c$ VEVs) in two different
$\lambda\! +\! \overline \lambda$ multiplets. The first
scale leaves a symmetry $SU(3)_C\times SU(2)_L\times
SU(2)_R\times U(1)_{B-L}$, which is subsequently
broken to the standard  $SU(3)_C\times SU(2)_L\times
U(1)_Y$ gauge group. Below the intermediate scales
the exotic flavors in the three chiral ${\bf 27}$s of $E_6$
(two down type quarks, two neutrinos, and two
lepton/Higgs doublets, see Eq.(3)) will
combine into nonchiral
representations of the standard model gauge group and
will become very massive.

\vskip 0.3truecm
\noindent 1.2. MATTER PARITY MODELS
\vskip 0.2truecm

To guarantee the absence of the lowest dimension
baryon and lepton number violating operators and then of
a fast proton decay, an effective matter parity must be at
work. It has been shown that only two $Z_N$ matter parities
can be implemented in the class of models under study [16].
They correspond to $P_2= C g_2$, generating a $Z_2$
discrete group, and $P_3= B g_3$, which defines a $Z_3$
symmetry (see Table I for the definition of $B$ and $C$).
The action of the gauge discrete symmetries $g_{2,3}$
on the {\bf 27} representation of $E_6$ and
the transformation properties of the standard matter fields
under $P_{2,3}$ are listed in Tables VI and VII, respectively.
Both of these matter parities
forbid in the low energy superpotential the terms
$$lh,\ lle^c,\ qld^c,\ u^cd^cd^c \eqno(6)$$
($l$, $h$, and $q$ stand for lepton,
Higgs and quark doublets)
while allowing for the standard Yukawa terms.
As $P_{2,3}$ are low energy symmetries, the VEVs breaking
the gauge group $G$ down to $SU(3)_C\times SU(2)_L\times U(1)_Y$
at large intermediate
scales must lie along matter parity neutral directions.
This implies that the $E_6$ singlets $s$ and the
$SU(5)$ singlets $\nu^c$ and $N$
acquiring VEVs must live in the subspaces generated by
the families in Table VIII.

\vskip 0.3truecm
\noindent 1.3. PATTERNS OF FERMION MASSES WITH TEXTURE ZEROES
\vskip 0.2truecm

The intermediate scales will break the discrete
symmetries of the model except for a low energy matter parity.
 Broken discrete symmetries, however,
may define texture zeroes in the fermion mass matrices [7].
Moreover, the existence of different scales of symmetry breaking
may generate the observed hierarchy of masses and mixing angles.
To illustrate this, consider a $Z_N\times Z_2$
symmetry acting on three chiral families of fermions $f_i$
as
$$\eqalign{(f_1,f_2,f_3)&\rightarrow
(\alpha^{-2} f_1,\alpha f_2,f_3)\ ,\cr
(f_1,f_2,f_3)&\rightarrow (f^c_1,f^c_2,f^c_3)\ ,\cr }
\eqno(7)$$
($\alpha^N=1$, with $N\not= 2,3,4,6$)
with analogous transformation properties for the antifermions
$f^c_i$.
Now suppose that the discrete symmetry is broken in the Higgs
sector, in such a way that the Higgs
$h=\kappa_1h_1+\kappa_2h_2+\kappa_3h_3$ contains components
in three families which transform under $Z_N$
$$\eqalign{(h_1,h_2,h_3)&\rightarrow
(h_1,\alpha^{-2} h_2,\alpha h_3)\ ,\cr
}\eqno(8)$$
and are neutral respect to the $Z_2$ symmetry.
When $h$ developes a VEV, the structure of the corresponding
fermion mass matrix will be
$$\bordermatrix{&&f_i&\cr
&\cdot&C&\cdot\cr
f^c_i&C&B&\cdot\cr
&\cdot&\cdot&A\cr}\ ,\eqno(9)$$
where $A$, $B$, and $C$ are proportional to
$\kappa_1$, $\kappa_2$, and $\kappa_3$, respectively.
Moreover, since the components $\kappa_i$ may be proportional
to ratios between intermediate scales (and/or $M_c$), the
hierarchy of these coefficients could explain
the observed pattern of fermion masses and mixings.
The texture above, for instance, in the down quark mass matrix,
could fit $m_b$, $m_s$, $m_d$ and predict (for a diagonal
{\it up--charm} submatrix) $V_{12}\approx \sqrt {m_d\over m_s}
\approx 0.22$.
Analogous mass matrix structures are found in the class of models
under study.

\vskip 0.3truecm
\centerline{\bf II. MODELS WITH $P_3$ MATTER PARITY}
\vskip 0.2truecm

In this section we classify the models with $P_3$
matter parity. We find that only three patterns of quark
mass matrices are consistent with the conditions {\it (i-iv)}
discussed
in section I. None of them, however, can accommodate the
observed values of quark masses and mixings.

To preserve $P_3$ (condition {\it (i)}) the fields developing
large VEVs must be combinations of the fields in Table VIII.
The 4 vectorlike families of quarks
$(u\ d)+\overline {(u\ d)}$ and
$u^c+\overline {u^c}$ will get masses (condition {\it (ii)})
via renormalizable
interactions only if one of the singlets $s_2,s_6,s_{14}$
and another of $s_4,s_8,s_{16}$ acquire VEVs.
To avoid fast proton decay
 it is also necessary
that the down type quarks $d'$ in $q_3$, $q_5$ and
$q_7$ become
very massive. This only happens if $N$ in $\lambda_2$,
$\lambda_4$ or $\lambda_5$ aquire VEVs ($\langle N\rangle$ in
$\lambda_6$ and
$\lambda_7$ leave $d'$ in $q_3$ massless).
A remarkable fact in $P_3$ models is that the matter parity
imposes the three chiral families of $(u\ d)$ and $u^c$
to be in $q_3$, $q_5$, $q_7$ and $Q_3$, $Q_5$, $Q_7$,
respectively. This implies that
below the intermediate scales the symmetries $A$,
$D\times V_d$, and $P\times V_d$ in Table I remain unbroken
in the up quark sector
and may generate matrix textures
{\it \`a la } Ramond-Roberts-Ross, as explained in the
Introduction.  These structures would be
approximate in the down sector.

According to the $P_3$ assignments
(note that the matter parity is an extra quantum number
that distinguishes lepton from Higgs doublets)
the standard model Higgses
$h$, $h'$
result from the diagonalization of the $9\times 9$ matrix
with
$h$ in $\lambda_2$, $\lambda_4$, $\lambda_5$,
$\lambda_6$, $\lambda_7$; $\overline{h'}$ in
$\overline \lambda_1$,
$\overline \lambda_2$; and $\overline l$ in
$\overline \lambda_5$, $\overline \lambda_6$
(rows), and
$h'$ in $\lambda_2$, $\lambda_4$, $\lambda_5$,
$\lambda_6$, $\lambda_7$; $\overline{h}$ in
$\overline \lambda_1$, $\overline \lambda_2$; and
$l$ in $\lambda_1$, $\lambda_8$ (columns).
For $\langle N\rangle$ in $\lambda_2$
and $\langle \nu^c\rangle$ in $\lambda_3$ or $\lambda_9$ we
obtain a
light Higgs pair ({\it i.e.}, a rank 8 Higgs mass matrix) for
singlet VEVs in $s_2,s_4$ or $s_2,s_8$ (condition {\it (iii)}).
In these cases the light Higgs $h$ lies in $\lambda_6$
(without mixing with other families)
and would predict patterns of up quark mass matrices antisymmetric
({\it type A}) to be discussed below.
 For singlet VEVs in
$s_2,s_8$;
$s_4,s_6$; or $s_6,s_8$ there are {\it two} light
Higgses after the intermediate scales (one of them along
$\lambda_6$). These cases, however, are not
consistent with the required flatness in the scalar potential
(condition {\it (iv)}).
For singlet VEVs in $s_2,s_8$,
for instance, it is impossible to adjust to zero simultaneously
the $N$ and $\nu^c$ $D$-terms and the $F$-terms
$F_{N_4}$, $F_{\nu^c_3}$, $F_{s_2}$, $F_{s_4}$, $F_{s_6}$,
$F_{s_{14}}$, and $F_{s_{16}}$.
The cases with  $\langle N \rangle$ along
$\lambda_4$ are analogous.

For $\langle N\rangle$ in $\lambda_5$
and $\langle \nu^c\rangle$ in $\lambda_3$
we obtain light
Higgses $h,h'$ for singlet VEVs in
$s_2,s_4$; or $s_2,s_8$.
The second case, however, does not take
place along a flat direction and is excluded. The first case
does define a flat direction, and gives the
pattern of quark masses and mixings of {\it type B}
(see below).
Finally, there are the models with  $\langle N\rangle$
in $\lambda_5$
and $\langle \nu^c\rangle$ in $\lambda_9$. The case with
singlet VEVs along $s_4,s_6$ does not lie along a flat
direction, while the model with VEVs along
$s_2,s_4$ gives the pattern of {\it type C} discussed below.

\vskip 0.3truecm
\noindent 2.1. QUARK MASS MATRICES IN $P_3$ MATTER PARITY
MODELS
\vskip 0.2truecm

\noindent $\bullet$ {\it Type A}: In this case the
 up quark mass matrices are generically antisymmetric as
a consequence of an exact symmetry. The
$\overline{(u\ d)}_a M_{ab} (u\ d)_b$ and
$\overline{u^c}_a M_{ab} u^c_b$ mass matrices defining
the three chiral families are identical, and then the
states which correspond to the
standard $(u\ d)$ and $u^c$ quarks lie in the same
directions in flavor space. Therefore, as under $P\times V_d$
the Higgs $h(\in \lambda_6)$ is odd whereas $u(\in q)$ and
$u^c(\in Q)$ transform into each other (see Table I), all
contributions to the up quark mass matrices are antisymmetric.
This
implies $m_t=m_c$ and $m_u=0$, making all these cases
unrealistic.

\noindent $\bullet$ {\it Type B}: The quark mass matrices in this
case are
$$\vbox{
\tabskip=19pt
\halign{#\hfil&#\hfil\cr
$\bordermatrix{
&&u_i&\cr
&\cdot&B&B'\cr
u^c_j&B&A'&\cdot\cr
&B'&\ \cdot\ &A\cr}\ ,$&
$\bordermatrix{
&&d_i&\cr
&D&\cdot&E\cr
d^c_j&F&\cdot&G\cr
&\cdot&C&\cdot\cr}\ ,$\cr
}}\eqno(10)$$
with $A/A'=B/B'$.
This structure can be disregarded only
after a detailed analysis.
It is possible to fit
the six quark masses, with a heavy {\it top} quark,
but the predicted
mixings of the third family are far too small. In
particular, we obtain $V_{23} < m_c^2/m_t^2\approx 10^{-4}$,
while experimentally this mixing is two orders of
magnitude larger ($V_{23}\approx 0.042\pm 0.12$ [17]).
The relatively large entries
with a nonperturbative origin required to cure this
problem would be in contradiction with our assumptions
(see a more detailed discussion below).

\noindent $\bullet$ {\it Type C}: This is the most
interesting $P_3$ case. The quark mass matrices read
$$\vbox{
\tabskip=19pt
\halign{#\hfil&#\hfil\cr
$\bordermatrix{
&&u_i&\cr
&\cdot&B&B'\cr
u^c_j&B&A'&\cdot\cr
&B'&\ \cdot\ &A\cr}\ ,$&
$\bordermatrix{
&&d_i&\cr
&C&\cdot&\cdot\cr
d^c_j&\cdot&D&E\cr
&\cdot&F&G\cr}\ ,$\cr
}}\eqno(11)$$
with $A/A'=B/B'$ but otherwise arbitrary
complex entries (since they depend on unknown normalization
factors of the fields).
The zeroes in these matrices are a remnant of the
discrete symmetry $A$ in Table I. For example,
for the up quarks we have:
$$\eqalign{(u_3,u_5,u_7)&\rightarrow
(u_3,\alpha u_5,\alpha u_7)\ ,\cr}
\eqno(12)$$
and for the Higgs ($h=\kappa_1 h_2+\kappa_2 h_4$):
$$\eqalign{(h_2,h_4)&\rightarrow
( \alpha^2 h_2,\alpha h_4)\ .\cr}
\eqno(13)$$
The up quark matrix is symmetric due to the
$P\times V_d$ symmetry, whereas
the relation between the coefficients follows from the
$D$ symmetry:
$$(u_3,u_5,u_7,h_2,h_4)\rightarrow
(u^c_3,u^c_7,u^c_5,h_4,h_2)\ .\eqno(14)$$
Although these structures are sensitive to
{\it top} radiative corrections, to decide on the main features
 of these matrices the following
approximate analysis will show up good enough. To avoid
degeneracy on quark masses, $A\approx m_t$ and $A'\approx m_c$
($A'\approx m_t$ and $A\approx m_c$ give equivalent
 results).
On the other hand,
since the first column of the down quark matrix
does not mix with
the other two, $B$ in the up quark matrix is fixed by the
Cabbibo angle ($\approx V_{12}$):
$B\approx A'V_{12}\approx m_c V_{12}$, implying
$B'\approx {m^2_c\over m_t}.$
In the down quark matrix
$C\approx m_d$ and the other nonzero entries give
$m_s$, $m_b$, and $V_{23}$. The mixing $V_{23}$ comes
mainly from the down sector because $B'\approx 0$,
which also implies
$V_{13}\approx V_{12} V_{23}$ and
$m_u\approx m_c V^2_{12}$. The last relation
gives a too high estimate of $m_u\approx$ 40 MeV.

The value of $m_u$ that we obtain seems the only
bad prediction of {\it type C} matrices, and we wonder if
contributions from nonrenormalizable terms could change it.
In order to fulfill conditions {\it (iii)} and {\it (iv)},
throughout the paper we have assumed a generic suppression
by 13 orders of magnitude for these terms and based our
analysis on trilinear (renormalizable) couplings.
To predict an acceptable value of $m_u$ higher order
contributions should introduce a direct {\it up} mass term or
a mixing term
between the first and second  down quark families.
These contributions would be of order
 $V^2_{12}m_c/m_t\approx 10^{-5}$ times the Yukawa
contribution corresponding to the
{\it top} quark mass, which seems to be
much higher than the suppression required by
conditions {\it (iii,iv)}. Similar but even more severe
comments apply to pattern {\it B} above. There the bad
prediction is $V_{23}<10^{-4}$, and to fix it one should rely
on nonrenormalizable contributions with a suppression
of order $10^{-2}$ relative to trilinear entries and
still keep the Higgs fields light and the flatness conditions.

\vskip 0.3truecm
\centerline{\bf III. MODELS WITH $P_2$ MATTER PARITY}
\vskip 0.2truecm

An analogous exploration can be done for models with the
$Z_2$ matter parity $P_2$. It is convenient to write the
superpotential in terms of $C$ eigenstates (see Table III),
which we denote with primes.
In $P_2$ models the Higgs $h,h'$
result from the diagonalization of a $11\times 11$ mass
matrix
where the rows correspond to
$h$ in $\lambda'_1$, $\lambda'_3$, $\lambda'_5$,
$\lambda'_6$, $\lambda'_8$; $\overline{h'}$ in
$\overline \lambda'_1$,
$\overline \lambda'_2$; and $\overline l$ in
$\overline \lambda'_3$, $\overline \lambda'_4$,
$\overline \lambda'_5$, $\overline \lambda'_6$,
and the columns to
$h'$ in $\lambda'_1$, $\lambda'_3$, $\lambda'_5$,
$\lambda'_6$, $\lambda'_8$; $\overline{h}$ in
$\overline \lambda'_1$, $\overline \lambda'_2$, and
$l$ in $\lambda'_2$, $\lambda'_4$,
$\lambda'_7$, $\lambda'_9$.
In $P_2$ models the
three chiral families of quarks will be
combinations of the families
$q'_1$, $q'_2$, $q'_3$, $q'_4$, and $q'_6$, with
all discrete symmetries (except for $P_2$) broken in the
quark sector.
Then, it is not obvious, but there are
restricted patterns of quark mass matrices in these models.

For singlet VEVs along $s'_1,s'_3,s'_{13},s'_{15}$
we obtain models which satisfy simultaneously
the flatness conditions while
making all vectorlike quarks heavy. In all these models
(for any choice of $N$ and $\nu^c$ VEVs),
however, the Higgs $h$ which couples to up quarks
lies in $\lambda'_6$ and/or in pure
$\overline {\lambda}'$ fields. The first case gives
antisymmetric matrices (pattern {\it type A} discussed above),
and the second has no Yukawa couplings and implies
massless up quarks.
Any other combination of VEVs, except for two equivalent
choices, do not satisfy conditions {\it (ii)} and {\it (iv)}
simultaneously. The only interesting case
is obtained when the singlet VEV has
components along $s'_1,s'_{7},s'_{13}$, whereas
$\langle N\rangle$
is along
$\lambda'_1$, $\langle \overline N\rangle$ along
$\overline \lambda'_{1,2}$,
$\langle \nu^c\rangle$ along $\lambda'_7$ and $\langle \overline
\nu^c\rangle$
along $\overline \lambda'_{3,4}$. (The
equivalent choice is obtained by a $D\times V_d$ transformation.)

Let us concentrate on this case. From now on we drop
the primes for specifying $C$ eigenstates.

\vskip 0.3truecm
\noindent 3.1. FLATNESS CONDITIONS
\vskip 0.2truecm

For the vacuum under consideration
$D$-flatness implies
$$\eqalign{\langle \nu^c_7\rangle=&\ \langle \overline {\nu^c}_3+
\overline {\nu^c}_4\rangle\ ,\cr
\langle N_1 \rangle =&
\ \langle \overline N_1+\overline N_2\rangle\  ,\cr}\eqno(15)$$
while $F$-flatness imposes
$$\eqalign{
\langle F_{\nu^c_4}\rangle=&\ \langle c_3\ s_7\overline \nu^c_3+
c_4\ s_7\overline \nu^c_4\rangle=0\ , \cr
\langle F_{N_3}\rangle=&\ \langle c_1\ s_1\overline N_1+
c_2\ s_1\overline N_2+c_9\ s_{13}\overline N_1+
c_{10}\ s_{13}\overline N_2\rangle=0\ ,\cr
\langle F_{s_3}\rangle=&\ \langle c_1\ N_1\overline N_1-
c_2\ N_1\overline N_2+d_7\ s^2_7\rangle=0\ ,\cr
\langle F_{s_{15}}\rangle=&\ \langle c_9\ N_1\overline N_1-
c_{10}\ N_1\overline N_2+d_8\ s^2_7\rangle=0\ ,\cr
\langle F_{s_1}\rangle
=&\ \langle 3d_1\ s^2_1+2d_2\ s_1s_{13}+d_4
\ s^2_{13}\rangle=0\ ,\cr
\langle F_{s_{13}}\rangle
=&\ \langle d_2\ s^2_1+3d_3\ s^2_{13}+2d_4
\ s_1s_{13}\rangle=0\ .\cr
}\eqno(16)$$
\vskip 0.2truecm
\noindent (The Yukawa couplings involving the gauge singlets
were defined in Table V.)
The rest of F-terms are trivially zero.
The 8 equations have to be solved
adjusting 9 VEVs. The 2 equations for $\langle \nu^c_7\rangle$,
$\langle \overline \nu^c_3\rangle$ and
$\langle \overline \nu^c_4\rangle$ decouple
from the rest, defining a flat direction.
The 4 homogeneous equations involving
$\langle N_1, \overline N_{1,2} \rangle $
and $\langle s_7 \rangle $ can be solved as a function
of $\langle {{s_{13}}\over {s_1}} \rangle $ and $c$
but after modifying
$\langle F_{s_3}\rangle$ and/or
$\langle F_{s_{15}}\rangle$, for otherwise the corresponding
equations are incompatible.
This is so because (we follow the procedure of
Ref.~[9] to calculate the Yukawa couplings)
$${{c_2}\over {c_1}} = {{c_{10}}\over {c_9}} =
\sqrt {3\over 2}\ ,
\eqno(17)$$
whereas $d_7$ and $d_8$ have a different $c$ dependence.
The simplest solution is to assume a large singlet mass
term $\ M\ s_{1,13}s_{3,15}\ $, with
$M\approx M_c$ if the VEVs have to be of the correct size.
Although
nonrenormalizable terms and singlet masses have both
nonperturbative origin,
due to the exponential behaviour of the scales generated
nonperturbatively [11] nothing prevents a very different
suppression for those two types of terms (as required here).
Finally, $\langle
F_{s_{1,13}}\rangle=0$ fix
$\langle {{s_{13}}\over {s_1}} \rangle $ and $c$
because they
are homogeneous and compatible only for a definite
choice of Yukawa couplings, and then of $c$.
The prove that the preferred vacuum alignment
is the one just discussed also
requires to know the soft scalar masses that trigger the
VEVs.

\vskip 0.3truecm
\noindent 3.2. QUARK MASS MATRICES IN THE $P_2$ MATTER PARITY
MODEL
\vskip 0.2truecm

Once fixed the pattern of symmetry breaking, we can study the
spectrum of fields that remain light and, in particular,
identify the three chiral families of quarks and leptons.
In Figs.~1-2 we give the
resulting mass matrices for lepton doublets ($1a$), charged
lepton singlets ($1b$), Higgs doublets ($1c$), down-quark
singlets ($2a$), and up-quark singlets and quark doublets
which coincide ($2b$).
We use $\langle N_1\rangle\sim  \langle\overline N_{1,2}\rangle
\sim x$,
 $\langle \nu^c_7\rangle\sim \langle\overline \nu^c_{3,4}\rangle
\sim y$,
$\langle s_{13}+s_{14}\rangle\sim r$, $\langle s_{7} \rangle\sim s$
to specify the order of magnitude of the entries.
A dot stands for a zero due to the
$P_2$ symmetry. Blanks may be
eventually filled out with
nonrenormalizable contributions of order 1 TeV.
Gaugino mass generation
has been taken into account introducing one pair of doublets
in $(2a)$ and two pairs of singlets in $(2b)$.
Diagonalizing these matrices we find the three
chiral families of quarks and leptons:

$$\vbox{
\tabskip=8pt
\halign{\hfil#&#\hfil&#\hfil&#\hfil\cr
${u\choose d}:\;\;$&$q_3$ ,&$\alpha_1q_1+\alpha_2
q_2+\alpha_3q_4$ ,&$\beta_1q_6+\beta_2
q_2+\beta_3q_4\ ;$\cr
${u^c}:\;\;$&${u^c}_3$ ,&
$\alpha_1{u^c}_1+\alpha_2{u^c}_2+\alpha_3{u^c}_4
$ ,&$\beta_1{u^c}_6+\beta_2{u^c}_2+\beta_3{u^c}_4\ ;$\cr
${d^c}:\;\;$&${d^c}_3$ ,&$\gamma_1{d^c}_1+
\gamma_2{d^c}_2+\gamma_3{d^c}_4+\gamma_4{d^c}_6$ ,
&$\delta_1{d'^c}_5+\delta_2{d^c}_2+\delta_3{d^c}_4
+\delta_4{d^c}_6\ ;$\cr}}\eqno(18)$$

$$\vbox{
\tabskip=8pt
\halign{#\hfil&#\hfil&#\hfil&#\hfil\cr
${e\choose \nu}:\;\;$&$\epsilon_1l_1+\epsilon_2h'_7
+\epsilon_3h'_9$ ,&$l_6$ ,&$l_8\ ;$\cr
${e^c}:\;\;$&${e^c}_5$ ,&${e^c}_6$ ,&${e^c}_8\ .$\cr}}
\eqno(19)$$

\noindent
There are two pairs of Higgs doublets
light after the intermediate scales (see Fig.~1c).
These Higgses receive
order 1 TeV masses only from nonrenormalizable (and soft SUSY
breaking) effects. We will assume that these masses mix them
and that the pair not involved in the electroweak
phase transition is massive enough ($\sim 1$ TeV) to avoid
flavour changing neutral currents via Yukawa interactions.
Then the Higgs components along {\bf 27} families are

$$\vbox{
\tabskip=8pt
\halign{\hfil#&#\hfil\cr
${h^0\choose h^+}:\;\;$&$\kappa_1h_6+
\kappa_2h_1\ ;$\cr
${h'^-\choose h'^0}:\;\;$&
$\kappa'_1h'_6+\kappa'_2h'_5+\kappa'_3h'_8+
\kappa'_4h'_3+\kappa'_5l_7\ .$\cr}}\eqno(20)$$

\noindent
All nonstandard lepton doublets and quarks
get heavy masses at the intermediate
scales. However, in addition to the extra pair of Higgs doublets,
one pair $(e^c,\overline e^c)$ of
charged lepton singlets remains light (see Fig.~1b), with
masses of order 1~TeV given by nonrenormalizable interactions.

When $h$ and $h'$ develope VEVs ($v$ and $v'$, respectively)
ordinary quarks and leptons get masses
through the Yukawa couplings in Table IV. To avoid a
Higgs along $h_6$ only ($\kappa_2=0$) and then an
antisymmetric {\it up} quark matrix, $s$ must be of the
same order as $x,y$. Now, at zero order
in $r/s$ $(\alpha_2=\alpha_3=0$, $\beta_2=\beta_3=0$,
$\gamma_i=\delta_i=0$ for $i\not= 1$) there are only two
entries in the up quark matrix:
$A\sim m_t$ and $B$, which would be much smaller than $A$
if $\kappa_1\ll \kappa_2$; and two entries in the
down quark matrix:
$A'\sim m_b$ and $B'$, verifying $B'\ll A'$ if
$\kappa'_2\ll \kappa'_3$. At zero
order in $\kappa'_2/\kappa'_3$, there are two entries in the
charged lepton
matrix: $A_1''$ and $A_2''$, which should satisfy
$A_1''^2+A_2''^2\sim m_\tau^2$. In consequence, at zero order in
$r/s$, $\kappa_2/\kappa_1$, and  $\kappa'_2/\kappa'_3$ only the
third family is massive. The structure in the
fermion matrices
will appear as a perturbation on these ratios.
The complete mass matrices are

\noindent $\bullet$ {\it Type D}:
$$\vbox{
\tabskip=19pt
\halign{#\hfil&#\hfil&#\hfil\cr
$\bordermatrix{
&&u_i&\cr
&\cdot&D&E\cr
u^c_j&D&\cdot&C\! +\! B\cr
&E&C\! -\! B&A\cr}\ ;$&
$\bordermatrix{
&&d_i&\cr
&B'&\cdot&\cdot\cr
d^c_j&\cdot&C'&A'\cr
&\cdot&D'&E'\cr}\ ;$&
$\bordermatrix{
&&e_i&\cr
&C''&B''&A_2''\cr
e^c_j&\cdot&D''&\cdot\cr
&E''&\cdot&A_1''\cr}\ ,$\cr
}}\eqno(21)$$
where $A$, $A'$, $A_1''$, and $A_2''$ are zero order entries
and the rest
correspond to higher order on the three ratios above.
All entries are arbitrary complex coefficients, with
the zeroes as a remnant of the $A\times B\times P\times V_d$
discrete symmetry
in Table I. (Although
$A$ and $B$
denote both discrete symmetries and mass matrix entries,
no confusion is expected by the use of this notation.)

All quark masses are taken at 1 GeV.
$m_t$, which we will define as $h_t/v$ at $M_Z$, must
be evolved down to 1 GeV. The QCD
{\it running} coefficient is a factor $\approx 1.8$.
Since $A\sim m'_t\equiv 1.8\ m_t$ and
in the down quark matrix the third family does not
participate of
the first column, $E\approx m_tV_{13}$
and $C^2-B^2\approx m_c m'_t$. The masses $m_d$ and $m_s$
and the mixing $V_{23}$
fix $B'$, $C'$, and $D'$: $B'\approx m_d$,
$C'\approx V_{23}m_b$ and
$D'\approx m_s$; whereas
for the Cabibbo mixing, $V_{12}$,
one has $D\approx m_c V_{12}$. Then $m_u$ results
from the cancellation
$$m_u\approx m_cV_{12}^2-1.8\ m_tV^2_{13}\ .\eqno(22)$$
This correlation translates into  a preference for a large
{\it top} quark mass and
$V_{13}$ mixing. For example, if $V_{13}\le 0.013$, the {\it
top} quark mass must be
 $m_t\ge 170$ GeV (for lower values of $V_{13}$ it is
necessary to include {\it top} radiative corrections
and a more precise diagonalization of the matrices to
give an estimate of the correlation).
The three charged lepton
masses can be easily  adjusted using the arbitrariness
in the third matrix.
A detailed numerical analysis of the range of variation
of the different (physical) parameters will be
presented elsewhere.

\vskip 0.3truecm
\noindent 3.3. OTHER PHENOMENOLOGICAL
IMPLICATIONS OF THE MODEL
\vskip 0.2truecm

Other aspect of the model are the phenomenological
implications of the nonstandard $(e^c,\overline e^c)$ pair
of leptons in the TeV region. They are even under $P_2$,
and then their only
couplings in the superpotential are
$$P= a\; lle^c +\; b\; h'h'e^c
+\; c\; hh\overline e^c +m\; e^c\overline e^c\ .
\eqno(23)$$
For reasonable values of their masses and couplings their
presence will not be in conflict with any particle physics
experiment. However, we note that these trilinears explicitly
break lepton number and could have
relevance in Cosmology (baryogenesis).

Finally, we consider the perturbative unification
of the gauge couplings in the model under study.
The extra pair of Higgs doublets and charged
lepton singlets, with masses around 1 TeV,
provides an evolution of the running
couplings which differs from the successful
evolution suggested in the minimal supersymmetric scenario
($\alpha_C=
\alpha_L=\alpha_Y$ at $M_X\approx 10^{16}$~GeV, see Fig.~3). In this
model, however, the presence of two intermediate
scales and a large number of vectorlike families below the
compactification scale seems to introduce enough arbitrariness
to obtain the right value of the electroweak mixing angle.
For $\sin^2_W=0.23$
the matter content implies that
$\alpha_L=\alpha_Y$ at $M\approx 10^{15}~GeV$ (see Fig.~3),
which sets the second intermediate scale $\langle \nu^c \rangle$.
Above this scale our model has left-right symmetry
({\it i.e.}, $\alpha_L=\alpha_R$). The unification of these
couplings with $\alpha_C$ at $M_c\ge 10^{16}~GeV$
requires the presence of nonstandard quarks below $M_c$.
Taking
three down quark singlets $d^c+\overline{d^c}$ with masses
around $10^{12}~GeV$ or 2 complete families of
$q+\overline q$, $Q+\overline Q$, $d'+d'^c$ at $10^{15}~GeV$
(the case plotted),
 and the rest of vectorlike families with a mass $M_c$, we
obtain
$\sin^2_W=0.23$ and a sensible unification
scale $M_c$ (see Fig.~3).

\vskip 0.3truecm
\centerline{\bf IV. SUMMARY AND CONCLUSIONS }
\vskip 0.2truecm

We have discussed all the possible
patterns of quark mass matrices in the
first and most extensively studied class of
3-generation
Calabi-Yau models. Our analysis is based on the discrete
symmetries of the models, and it includes the complete
matter content ($E_6$-singlet and -nonsinglet fields) and
their trilinear couplings
at the compactification scale.
 To select the {\it realistic} cases,
we have required an
unbroken low-energy matter parity (a $Z_2$ or a $Z_3$
discrete symmetry), absence of light leptoquarks
(which would be inconsistent with the proton lifetime
and perturbative unification), and intermediate scales
defining flat directions at the renormalizable
level. The models with a $P_3$ matter parity
left give 3 patterns of quark
mass matrices; none of them, however, is able to
accommodate a realistic spectrum of masses and mixing
angles. Among the models with a $P_2$ matter parity
we find only one case (studied in detail in sections 3.1-3.3)
 which seems free of inconsistencies, predicting an acceptable
pattern of fermion masses and mixings, with
preference for a heavy {\it top} quark ($m_t\ge 170$ GeV
for $V_{13}\le 0.013$).

The model we have singled out can be derived from the
string and is realistic. However, it can be seen as a
counterexample of the present searches for realistic
low-energy supersymmetric models. These are mainly based on two
observations: that perturbative unification
in the minimal model  predicts
the electroweak mixing angle with great accuracy and
that the observed pattern of fermion masses and mixing angles
can be explained with simple matrix structures (symmetric and
with texture zeroes). The model found is nonminimal, has
abundant extra matter near the unification scale, and has two
large intermediate scales of symmetry breaking.
In addition, although the fermion mass matrices have texture
zeroes and a perturbative structure which is  remnant of broken
discrete symmetries, these are not symmetric nor simple.

\vskip 0.3truecm
\noindent{\bf Acknowledgments}
\vskip 0.2truecm

We thank G.G. Ross for enlightening discussions.

\vskip 0.3truecm
\noindent{\bf References}
\vskip 0.2truecm

\noindent [1]
For a review on grand unification models, see
G.G. Ross, {\it Grand Unification Theories} (Frontiers in Physics,
Redwood City, CA, 1984).

\noindent [2]
For a review on string theory, see
M. Green, J. Schwarz and E. Witten, {\it Superstring Theory}
(Cambridge University Press, Cambridge, 1987).

\noindent [3]
P. Candelas, G. Horowitz, A Strominger, and E. Witten,
Nucl. Phys. {\bf B258} (1985) 75.

\noindent [4]
S.T. Yau, in {\it Simposium on Anomalies, Geometry and
Topology}, ed. W.A. Bardeen and A.R. White (World Scientific,
Singapore, 1985).

\noindent [5]
B.R. Green, K.H. Kirklin, P. Miron and G.G. Ross,
Nucl. Phys. {\bf B278} (1986) 667;
Nucl. Phys. {\bf B292} (1987) 606;
G.G. Ross, lectures given at 1988 {\it Banf Summer
Institute}, CERN preprint TH-5109/88.

\noindent [6] S. Dimopoulos, L.J. Hall, and S. Raby,
Phys. Rev. Lett. {\bf 68} (1992) 1984; J. Distler,
Phys. Rev. {\bf D45} (1992) 4195;
H. Arason, D.J. Casta\~no, P. Ramond, and E.J. Piard,
Phys. Rev.  {\bf D47} (1993) 1232.

\noindent [7]
P. Ramond, R.G. Roberts, and G.G. Ross,
Nucl. Phys. {\bf B406} (1993) 19; see also L. Ib\'a\~nez,
and G.G. Ross, Oxford preprint OUTP-9403; FTUAM94/7.

\noindent [8] B.R. Green, K.H. Kirklin, P. Miron and G.G. Ross,
Phys. Lett. {\bf B192} (1987) 111; J. Distler,
B.R. Green, K.H. Kirklin, P. Miron and G.G. Ross,
Phys. Lett. {\bf B195} (1987) 41.

\noindent [9]
M.G. Eastwood and T. Hubsch, Commun. Math. Phys. {\bf 132}
(1990) 383;
P. Berglund, L. Parkes, and T. Hubsch,
Commun. Math. Phys. {\bf 148} (1992) 57;
P. Berglund, T. Hubsch, and L. Parkes,
Mod. Phys. Lett. {\bf A5} (1990) 1485.

\noindent [10]
H.P. Nilles, Phys. Lett. {\bf 115B} (1982) 193; M. Dine,
R. Rohm, N. Seiberg, and E. Witten, Phys. Lett. {\bf 156B}
(1985) 55.

\noindent [11]
M. Dine, N. Seiberg, X. Wen, and E. Witten,
Nucl. Phys. {\bf B278} (1986) 769;
Nucl. Phys. {\bf B289} (1987) 319.

\noindent [12]
 F. del Aguila, G.D. Coughlan, and M. Masip, Phys.
Lett. {\bf 227B} (1989) 55.

\noindent [13]
G. Lazarides, P.K. Mohapatra, C. Panagiotakopoulos and
Q. Shafi, Nucl. Phys. {\bf B233} (1989) 614;
Berglund and T. Hubsch, Phys. Lett. {\bf 260B} (1991) 32.

\noindent [14]
G. Lazarides, C. Panagiotakopoulos, and
Q. Shafi, Phys. Lett. {\bf 225B} (1989) 66.

\noindent [15]
F. del Aguila, G.D. Coughlan, and M. Masip,
Nucl. Phys. {\bf B351} (1991) 90.

\noindent [16]
M.C. Bento, L. Hall and G.G. Ross, Nucl. Phys. {\bf B292}
(1987) 400; G. Lazarides and Q. Shafi,
Nucl. Phys. {\bf B338} (1990) 442.

\noindent [17]
{\it Review of Particle Properties},
M. Aguilar Ben\'\i tez {\it et al.},
Phys. Rev. {\bf D45} (1992) 1.

\vfill\eject

{\bf Table captions}

\noindent{\bf Table I:} Tensor representatives and
transformation properties of the
nonsinglet fields under the generators of the group
of discrete symmetries. The
transformations for the
$Q$ and $\overline Q$ multiplets follow from
interchanging $q\leftrightarrow Q$ and $\overline q
\leftrightarrow \overline Q$. ($\alpha^3=1.$)

\noindent{\bf Table II:} Tensor representatives and
transformation properties of the
$E_6$-singlet fields under the generators of the group
of discrete symmetries.

\noindent{\bf Table III:} $C$-eigenstates (primed) in terms of
$B$-eigenstates. $B$ and $C$ are defined in Tables I,II.

\noindent{\bf Table IV:} Trilinear couplings of type
$\lambda^3$ and $\lambda q Q$ allowed by the discrete
symmetries of the compactified model. We specify the terms
in the $B$ and the $C$ bases.

\noindent{\bf Table V:} Trilinear couplings of type
$s \lambda\overline \lambda $ and $s^3$
(in the $C$ basis) allowed by the discrete
symmetries.

\noindent{\bf Table VI:} Transformation properties of the
flavors in a {\bf 27} irrep of $E_6$ under the discrete
gauge symmetries $g_2$ and $g_3$. ($\alpha^3=1.$)

\noindent{\bf Table VII:} Transformation properties of
the standard quark, lepton, and Higgs superfields under
the matter parities $P_2$ and $P_3$.

\noindent{\bf Table VIII:} Neutral flavors under the
discrete symmetries $P_2$ and $P_3$. Large VEVs along
these directions leave unbroken the corresponding low
energy matter parity.

\vskip 0.3truecm

{\bf Figure captions}

\noindent{\bf Figure 1:} Mass matrices for lepton doublets
(a), charged lepton singlets (b), and Higgs doublets
(c).

\noindent{\bf Figure 2:} Mass matrices for down quark singlets
(a) and quark doublets and up quark singlets (b),
which coincide.

\noindent{\bf Figure 3:} Unification of
running coupling constants
for $\sin^2\theta_W=0.23$ in the
supersymmetric model with minimal
matter content (dashes) and
nonminimal supersymmetric model with an
extra $h+h'$ and $e^c+\overline e^c$ at 1 TeV (solid).
This model implies
left-right unification at a scale
$x\approx 10^{15}$ GeV. At this scale we add two families of
$q+\overline q$, $Q+\overline Q$, and $d'+d'^c$.

\vfill\eject\end

*****************************************************************

\nopagenumbers

\baselineskip=18pt
$$
\vbox{
\tabskip=1.truecm
\halign{#&#&\hfil#\hfil&\hfil#\hfil&\hfil#\hfil&\hfil#\hfil&\hfil#\hfil\cr
\noalign{\hrule}
\omit&\omit&\omit&\omit&\omit\cr
Field&Tensor&A&B&C&$D\times V_d$&$P\times V_d$\cr
\omit&\omit&\omit&\omit&\omit&\omit&\omit\cr
\noalign{\hrule}
\omit&\omit&\omit&\omit&\omit&\omit&\omit\cr
$\lambda_1$&$\phi_{(123)}$&$\alpha^2$
&$\alpha$&$\lambda_2$&$
\lambda_3$&1\cr
$\lambda_2$&$\phi_{(124)}$&$\alpha^2$&
$1$&$\lambda_1$&$
\lambda_4$&1\cr
$\lambda_3$&$\phi_{(\underline 1\underline 2\underline 3)}$
&$\alpha$&$\alpha^2$&$\lambda_4$&$
\lambda_1$&1\cr
$\lambda_4$&$\phi_{(\underline 1\underline 2\underline 4)}$
&$\alpha$&$1$&$\lambda_3$&$
\lambda_2$&1\cr
$\lambda_{5}$&$\phi_{3\underline 3}+\phi_{4\underline 4}$
&1&1&1&1&1\cr
$\lambda_{6}$&$\phi_{2\underline 2}-\phi_{1\underline 1}$
&1&1&1&1&$-1$\cr
$\lambda_{7}$&$\phi_{3\underline 3}-\phi_{4\underline 4}$
&1&1&$-1$&1&1\cr
$\lambda_8$&$\phi_{3\underline 4}$
&$1$&$\alpha$&$\lambda_9$&$
\lambda_9$&1\cr
$\lambda_9$&$\phi_{4\underline 3}$
&1&$\alpha^2$&$\lambda_8$&$
\lambda_8$&1\cr
\omit&\omit&\omit&\omit&\omit\cr
\noalign{\hrule}
\omit&\omit&\omit&\omit&\omit\cr
$\overline{\lambda_1}$&$(J_x+J_y)/\sqrt 3$&1&1&1&1&1\cr
$\overline{\lambda_2}$&$(J_x-J_y)/\sqrt 2$&1&1&1&$-1$&1\cr
$\overline{\lambda_3}$&$\epsilon \phi^{(34)}$
&$\alpha$&$\alpha$&$-1$&$
\overline{\lambda_6}$&$-1$\cr
$\overline{\lambda_4}$&$\underline \epsilon \phi^{(\underline 1
\underline 2)}$
&$\alpha$&$\alpha$&$-1$&$
\overline{\lambda_5}$&$-1$\cr
$\overline{\lambda_5}$&$\epsilon \phi^{(12)}$
&$\alpha^2$&$\alpha^2$&$-1$&$
\overline{\lambda_4}$&$-1$\cr
$\overline{\lambda_6}$&$\underline \epsilon \phi^{(\underline 3
\underline 4)}$
&$\alpha^2$&$\alpha^2$&$-1$&$
\overline{\lambda_3}$&$-1$\cr
\omit&\omit&\omit&\omit&\omit&\omit&\omit\cr
\noalign{\hrule}
\omit&\omit&\omit&\omit&\omit&\omit&\omit\cr
$q_1$&$\phi_{(234)}$
&$\alpha$&$\alpha$&1&$ Q_2$&$ Q_1$\cr
$q_2$&$\phi_{(\underline 1\underline 3\underline 4)}$
&$\alpha^2$&$\alpha^2$&1&$ Q_1$&$ Q_2$\cr
$q_3$&$\phi_{1\underline 2}$
&1&$1$&1&$ Q_3$&$ Q_3$\cr
$q_4$&$\phi_{2\underline 3}$
&$\alpha$&$\alpha^2$&$ q_5$&$
Q_6$&$ Q_4$\cr
$q_5$&$\phi_{2\underline 4}$
&$\alpha$&$1$&$ q_4$&$
Q_7$&$ Q_5$\cr
$q_6$&$\phi_{3\underline 1}$
&$\alpha^2$&$\alpha$&$ q_7$&$
Q_4$&$ Q_6$\cr
$q_7$&$\phi_{4\underline 1}$
&$\alpha^2$&$1$&$ q_6$&$
Q_5$&$Q_7$\cr
\omit&\omit&\omit&\omit&\omit\cr
\noalign{\hrule}
\omit&\omit&\omit&\omit&\omit\cr
$\overline{q_1}$&$\epsilon \phi^{(23)}$
&1&$\alpha$&$
\overline{q_2}$&$
\overline{Q_4}$&$ \overline{Q_1}$\cr
$\overline{q_2}$&$-\epsilon \phi^{(24)}$
&1&$\alpha^2$&$
\overline{q_1}$&$
\overline{Q_3}$&$ \overline{Q_2}$\cr
$\overline{q_3}$&$\underline\epsilon\phi^{(\underline 1\underline 4)}$
&1&$\alpha$&$
\overline{q_4}$&$
\overline{Q_2}$&$
\overline{Q_3}$\cr
$\overline{q_4}$&$-\underline\epsilon\phi^{(\underline 1\underline 3)}$
&1&$\alpha^2$&$
\overline{q_3}$&$
\overline{Q_1}$&$ \overline{Q_4}$\cr
\omit&\omit&\omit&\omit&\omit&\omit&\omit\cr
\noalign{\hrule}
}}$$
\centerline{{\bf Table I}}

\vfill
\eject

\baselineskip=18pt
$$
\vbox{
\tabskip=1.truecm
\halign{#&#&\hfil#\hfil&\hfil#\hfil&\hfil#\hfil&\hfil#\hfil&\hfil#\hfil\cr
\noalign{\hrule}
\omit&\omit&\omit&\omit&\omit\cr
Field&Tensor&A&B&C&$D$&$P$\cr
\omit&\omit&\omit&\omit&\omit&\omit&\omit\cr
\noalign{\hrule}
\omit&\omit&\omit&\omit&\omit&\omit&\omit\cr
$s_1$&$\phi_{1(23)}+\phi_{2(13)}$&$\alpha^2$&$\alpha$&$s_2$&$s_3$&$1$\cr
$s_2$&$\phi_{1(24)}+\phi_{2(14)}$&$\alpha^2$&$1$&$s_1$&$s_4$&$1$\cr
$s_3$&$\phi_{\underline 1 (\underline 2 \underline 3)}+
\phi_{\underline 2 (\underline 1 \underline 3)}$&
$\alpha$&$\alpha^2$&$s_4$&$s_1$&$1$\cr
$s_4$&$\phi_{\underline 1 (\underline 2 \underline 4)}+
\phi_{\underline 2 (\underline 1 \underline 4)}$&
$\alpha$&$1$&$s_3$&$s_2$&$1$\cr
$s_5$&$\phi_{1(23)}-\phi_{2(13)}$&$\alpha^2$&$\alpha$&$s_6$&$s_7$&$-1$\cr
$s_6$&$\phi_{1(24)}-\phi_{2(14)}$&$\alpha^2$&$1$&$s_5$&$s_8$&$-1$\cr
$s_7$&$\phi_{\underline 1 (\underline 2 \underline 3)}-
\phi_{\underline 2 (\underline 1 \underline 3)}$&
$\alpha$&$\alpha^2$&$s_8$&$s_5$&$-1$\cr
$s_8$&$\phi_{\underline 1 (\underline 2 \underline 4)}-
\phi_{\underline 2 (\underline 1 \underline 4)}$&
$\alpha$&$1$&$s_7$&$s_6$&$-1$\cr
$s_{9}$&$\phi_{3(44)}-\phi_{4(34)}$&$1$&$\alpha$&$s_{10}$&$s_{11}$&$1$\cr
$s_{10}$&$\phi_{4(33)}-\phi_{3(43)}$&$1$&$\alpha^2$&$s_9$&$s_{12}$&$1$\cr
$s_{11}$&$\phi_{\underline 3 (\underline 4 \underline 4)}-
\phi_{\underline 4 (\underline 3 \underline 4)}$&
$1$&$\alpha^2$&$s_{12}$&$s_9$&$1$\cr
$s_{12}$&$\phi_{\underline 4 (\underline 3 \underline 3)}
-\phi_{\underline 3 (\underline 4 \underline 3)}$&
$1$&$\alpha$&$s_{11}$&$s_{10}$&$1$\cr
$s_{13}$&$\epsilon \underline \epsilon
\phi^{(\underline 1\underline 2\underline 3)}$&
$\alpha^2$&$\alpha$&$s_{14}$&$s_{15}$&$1$\cr
$s_{14}$&$\epsilon \underline \epsilon
\phi^{(\underline 1\underline 2\underline 4)}$&
$\alpha^2$&$1$&$s_{13}$&$s_{16}$&$1$\cr
$s_{15}$&$\epsilon \underline \epsilon
\phi^{( 1  2  3)}$&
$\alpha$&$\alpha^2$&$s_{16}$&$s_{13}$&$1$\cr
$s_{16}$&$\epsilon \underline \epsilon
\phi^{( 1  2  4)}$&
$\alpha$&$1$&$s_{15}$&$s_{14}$&$1$\cr
$s_{17}$&$\epsilon
\phi^{(12)}$&$\alpha^2$&$\alpha^2$&$-1$&$s_{18}$&$-1$\cr
$s_{18}$&$\underline \epsilon
\phi^{(\underline 1 \underline 2)}$&
$\alpha$&$\alpha$&$-1$&$s_{17}$&$-1$\cr
$s_{19}$&$\epsilon
\phi^{(34)}$&$\alpha$&$\alpha$&$-1$&$s_{20}$&$-1$\cr
$s_{20}$&$\underline \epsilon
\phi^{(\underline 3 \underline 4)}$&
$\alpha^2$&$\alpha^2$&$-1$&$s_{19}$&$-1$\cr
\omit&\omit&\omit&\omit&\omit&\omit&\omit\cr
\noalign{\hrule}
}}$$
\centerline{{\bf Table II}}

\vfill
\eject

$$
\vbox{
\tabskip=1.5truecm
\halign{#\hfil&#\hfil&#\hfil\cr
\noalign{\hrule}
\omit&\omit&\omit \cr
Leptons&Quarks &Singlets\cr
\omit&\omit&\omit \cr
\noalign{\hrule}
\omit&\omit&\omit \cr
$\lambda'_1={1\over \sqrt 2}(\lambda_1+\lambda_2)$&
$q'_1=q_1$&$s'_1={1\over \sqrt 2}(s_1+s_2)$\cr
$\lambda'_2={1\over \sqrt 2}(\lambda_1-\lambda_2)$&
$q'_2=q_2$&$s'_2={1\over \sqrt 2}(s_1-s_2)$\cr
$\lambda'_3={1\over \sqrt 2}(\lambda_3+\lambda_4)$&
$q'_3=q_3$&$s'_3={1\over \sqrt 2}(s_3+s_4)$\cr
$\lambda'_4={1\over \sqrt 2}(\lambda_3-\lambda_4)$&
$q'_4={1\over \sqrt 2}(q_4+q_5)$&$s'_4={1\over \sqrt 2}(s_3-s_4)$\cr
$\lambda'_5=\lambda_5$&
$q'_5={1\over \sqrt 2}(q_4-q_5)$&$s'_5={1\over \sqrt 2}(s_5+s_6)$\cr
$\lambda'_6=\lambda_6$&
$q'_6={1\over \sqrt 2}(q_6+q_7)$&$s'_6={1\over \sqrt 2}(s_5-s_6)$\cr
$\lambda'_7=\lambda_7$&
$q'_7={1\over \sqrt 2}(q_6-q_7)$&$s'_7={1\over \sqrt 2}(s_7+s_8)$\cr
$\lambda'_8={1\over \sqrt 2}(\lambda_8+\lambda_9)$&
&$s'_8={1\over \sqrt 2}(s_7-s_8)$\cr
$\lambda'_9={1\over \sqrt 2}(\lambda_8-\lambda_9)$&
$\overline q'_1={1\over \sqrt 2}(\overline q_1+\overline q_2)$&
$s'_9={1\over \sqrt 2}(s_{9}+s_{10})$\cr
&$\overline q'_2={1\over \sqrt 2}(\overline q_1-\overline q_2)$&
$s'_{10}={1\over \sqrt 2}(s_{9}-s_{10})$\cr
$\overline \lambda'_1=\overline \lambda_1 $&
$\overline q'_3={1\over \sqrt 2}(\overline q_3+\overline q_4)$&
$s'_{11}={1\over \sqrt 2}(s_{11}+s_{12})$\cr
$\overline \lambda'_2=\overline \lambda_2 $&
$\overline q'_4={1\over \sqrt 2}(\overline q_3-\overline q_4)$&
$s'_{12}={1\over \sqrt 2}(s_{11}-s_{12})$\cr
$\overline \lambda'_3=\overline \lambda_3 $&&
$s'_{13}={1\over \sqrt 2}(s_{13}+s_{14})$\cr
$\overline \lambda'_4=\overline \lambda_4 $&&
$s'_{14}={1\over \sqrt 2}(s_{13}-s_{14})$\cr
$\overline \lambda'_5=\overline \lambda_5 $&&
$s'_{15}={1\over \sqrt 2}(s_{15}+s_{16})$\cr
$\overline \lambda'_6=\overline \lambda_6 $&&
$s'_{16}={1\over \sqrt 2}(s_{15}-s_{16})$\cr
&&$s'_{17}={1\over \sqrt 2}(s_{17}+s_{18})$\cr
&&$s'_{18}={1\over \sqrt 2}(s_{17}-s_{18})$\cr
&&$s'_{19}={1\over \sqrt 2}(s_{19}+s_{20})$\cr
&&$s'_{20}={1\over \sqrt 2}(s_{19}-s_{20})$\cr
\omit&\omit&\omit \cr
\noalign{\hrule}
\omit&\omit&\omit \cr}}$$
\centerline{{\bf Table III}}

\vfill\eject

\baselineskip=18pt

$$
\vbox{
\tabskip=1.truecm
\halign{\hfil#\hfil&#\hfil\cr
\noalign{\hrule}
\omit&\omit \cr
Coupling&Term\cr
\omit&\omit \cr
\noalign{\hrule}
\omit&\omit \cr
$a_{1}$&$\lambda_{1}\lambda_{4}\lambda_{9}+
\lambda_{2}\lambda_{3}\lambda_{8}={1\over \sqrt 2} (
\lambda'_{1}\lambda'_{3}\lambda'_{8}+
\lambda'_{1}\lambda'_{4}\lambda'_{9}-
\lambda'_{2}\lambda'_{3}\lambda'_{9}-
\lambda'_{2}\lambda'_{4}\lambda'_{8})$\cr
$a_{2}$&$\lambda^3_{1}+\lambda^3_{2}+\lambda^3_{3}+\lambda^3_{4}
={1\over \sqrt 2} (\lambda'^3_{1}+\lambda'^3_{3}+
3\lambda'_1\lambda'^2_2+3\lambda'_3\lambda'^2_4)$\cr
$a_{3}$&$\lambda^3_{8}+\lambda^3_{9}={1\over \sqrt 2} (
\lambda'^3_{8}+3\lambda'_8\lambda'^2_{9})$\cr
$a_{4}$&$\lambda_{5}\lambda_{1}\lambda_{3}+
\lambda_{5}\lambda_{2}\lambda_{4}=
\lambda'_{5}\lambda'_{1}\lambda'_{3}+
\lambda'_{5}\lambda'_{2}\lambda'_{4}$\cr
$a_{5}$&$\lambda_{7}\lambda_{1}\lambda_{3}-
\lambda_{7}\lambda_{2}\lambda_{4}=
\lambda'_{7}\lambda'_{1}\lambda'_{4}+
\lambda'_{7}\lambda'_{2}\lambda'_{3}$\cr
$a_{6}$&$\lambda_{5}\lambda_{8}\lambda_{9}={1\over 2} (
\lambda'_{5}\lambda'^2_{8}-\lambda'_{5}\lambda'^2_{9})$\cr
$a_{7}$&$\lambda_{5}\lambda^2_{6}=\lambda'_{5}\lambda'^2_{6}$\cr
$a_{8}$&$\lambda^3_{5}=\lambda'^3_{5}$\cr
$a_{9}$&$\lambda_{5}\lambda^2_{7}=\lambda'_{5}\lambda'^2_{7}$\cr
\omit&\omit \cr
\noalign{\hrule}
\omit&\omit \cr
$b_{1}$&$\lambda_{1} q_{2} Q_{7}+\lambda_{2} q_{2} Q_{6}+
\lambda_{3} q_{5} Q_{1}+\lambda_{4} q_{4} Q_{1}+ (q\leftrightarrow Q)$\cr
&$=\lambda'_{1} q'_{2} Q'_{6}-\lambda'_{2} q'_{2} Q'_{7}+
\lambda'_{3} q'_{4} Q'_{1}+\lambda'_{4} q'_{5} Q'_{1}+
(q'\leftrightarrow Q')$\cr
$b_{2}$&$\lambda_{1} q_{3} Q_{4}+\lambda_{2} q_{3} Q_{5}+
\lambda_{3} q_{6} Q_{3}+\lambda_{4} q_{7} Q_{3}+ (q\leftrightarrow Q)$\cr
&$=\lambda'_{1} q'_{3} Q'_{4}+\lambda'_{2} q'_{3} Q'_{5}+
\lambda'_{3} q'_{6} Q'_{3}+\lambda'_{4} q'_{7} Q'_{3}+
(q'\leftrightarrow Q')$\cr
$b_{3}$&$\lambda_{1} q_{6} Q_{6}+\lambda_{2} q_{7} Q_{7}+
\lambda_{3} q_{4} Q_{4}+\lambda_{4} q_{5} Q_{5}$\cr
&$={1\over \sqrt 2} (\lambda'_{1} q'_{6} Q'_{6}+\lambda'_{1} q'_{7} Q'_{7}+
\lambda'_{2} q'_{6} Q'_{7}+\lambda'_{2} q'_{7} Q'_{6}+$\cr
&$\;\;\;\lambda'_{3} q'_{4} Q'_{4}+\lambda'_{3} q'_{5} Q'_{5}+
\lambda'_{4} q'_{4} Q'_{5}+\lambda'_{4} q'_{5} Q'_{4})$\cr
$b_{4}$&$\lambda_{8} q_{1} Q_{6}+\lambda_{8} q_{5} Q_{2}+
\lambda_{9} q_{4} Q_{2}+\lambda_{9} q_{1} Q_{7}+ (q\leftrightarrow Q)$\cr
&$=\lambda'_{8} q'_{1} Q'_{6}+\lambda'_{8} q'_{4} Q'_{2}-
\lambda'_{9} q'_{5} Q'_{2}+
\lambda'_{9} q'_{1} Q_{7}+ (q'\leftrightarrow Q')$\cr
$b_{5}$&$\lambda_{8} q_{4} Q_{7}+\lambda_{9} q_{5} Q_{6}+
(q\leftrightarrow Q)$\cr
&$={1\over \sqrt 2} [\lambda'_{8} q'_{4} Q'_{6}-\lambda'_{8} q'_{5} Q'_{7}-
\lambda'_{9} q'_{4} Q'_{7}+\lambda'_{9} q'_{5} Q'_{6}+
(q'\leftrightarrow Q')]$\cr
$b_{6}$&$\lambda_{5} q_{1} Q_{2}+\lambda_{5} q_{2} Q_{1}
=\lambda'_{5} q'_{1} Q'_{2}+\lambda'_{5} q'_{2} Q'_{1}$\cr
$b_{7}$&$\lambda_{6} q_{1} Q_{2}-\lambda_{6} q_{2} Q_{1}=
\lambda'_{6} q'_{1} Q'_{2}-\lambda'_{6} q'_{2} Q'_{1}$\cr
$b_{8}$&$\lambda_{5} q_{4} Q_{6}+\lambda_{5} q_{5} Q_{7}
+ (q\leftrightarrow Q)=
\lambda'_{5} q'_{4} Q'_{6}+\lambda'_{5} q'_{5} Q'_{7}
+ (q'\leftrightarrow Q')$\cr
$b_{9}$&$\lambda_{6} q_{4} Q_{6}+\lambda_{6} q_{5} Q_{7}
- (q\leftrightarrow Q)=
\lambda'_{6} q'_{4} Q'_{6}+\lambda'_{6} q'_{5} Q'_{7}
- (q'\leftrightarrow Q')$\cr
$b_{10}$&$\lambda_{5} q_{3} Q_{3}=\lambda'_{5} q'_{3} Q'_{3}$\cr
$b_{11}$&$\lambda_{7} q_{4} Q_{6}-\lambda_{7} q_{5} Q_{7}
+ (q\leftrightarrow Q)=
\lambda'_{7} q'_{4} Q'_{7}+\lambda'_{7} q'_{5} Q'_{6}
+ (q'\leftrightarrow Q')$\cr
\omit&\omit \cr
\noalign{\hrule}
\omit&\omit \cr}}$$
\centerline{{\bf Table IV}}

\vfill\eject

$$
\vbox{
\tabskip=0.7truecm
\halign{\hfil#\hfil&#\hfil\cr
\noalign{\hrule}
\omit&\omit \cr
Coupling&Term\cr
\omit&\omit \cr
\noalign{\hrule}
\omit&\omit \cr
$c_{1}$&$s'_{1} \lambda'_{3} \overline \lambda'_{1}+
s'_{2} \lambda'_{4} \overline \lambda'_{1}+
s'_{3} \lambda'_{1} \overline \lambda'_{1}+
s'_{4} \lambda'_{2} \overline \lambda'_{1}$\cr
$c_{2}$&$s'_{1} \lambda'_{3} \overline \lambda'_{2}+
s'_{2} \lambda'_{4} \overline \lambda'_{2}-
s'_{3} \lambda'_{1} \overline \lambda'_{2}-
s'_{4} \lambda'_{2} \overline \lambda'_{2}$\cr
$c_{3}$&$s'_{5} \lambda'_{2} \overline \lambda'_{6}-
s'_{6} \lambda'_{1} \overline \lambda'_{6}+
s'_{7} \lambda'_{4} \overline \lambda'_{3}-
s'_{8} \lambda'_{3} \overline \lambda'_{3}$\cr
$c_{4}$&$s'_{5} \lambda'_{2} \overline \lambda'_{5}-
s'_{6} \lambda'_{1} \overline \lambda'_{5}+
s'_{7} \lambda'_{4} \overline \lambda'_{4}-
s'_{8} \lambda'_{3} \overline \lambda'_{4}$\cr
$c_{5}$&$s'_{5} \lambda'_{8} \overline \lambda'_{3}-
s'_{6} \lambda'_{9} \overline \lambda'_{3}+
s'_{7} \lambda'_{9} \overline \lambda'_{6}-
s'_{8} \lambda'_{8} \overline \lambda'_{6}$\cr
$c_{6}$&$s'_{5} \lambda'_{8} \overline \lambda'_{4}-
s'_{6} \lambda'_{9} \overline \lambda'_{4}+
s'_{7} \lambda'_{9} \overline \lambda'_{5}-
s'_{8} \lambda'_{8} \overline \lambda'_{5}$\cr
$c_{7}$&$s'_{9} \lambda'_{8} \overline \lambda'_{1}-
s'_{10} \lambda'_{9} \overline \lambda'_{1}+
s'_{11} \lambda'_{8} \overline \lambda'_{1}-
s'_{12} \lambda'_{9} \overline \lambda'_{1}$\cr
$c_{8}$&$s'_{9} \lambda'_{8} \overline \lambda'_{2}-
s'_{10} \lambda'_{9} \overline \lambda'_{2}-
s'_{11} \lambda'_{8} \overline \lambda'_{2}+
s'_{12} \lambda'_{9} \overline \lambda'_{2}$\cr
$c_{9}$&$s'_{13} \lambda'_{3} \overline \lambda'_{1}+
s'_{14} \lambda'_{4} \overline \lambda'_{1}+
s'_{15} \lambda'_{1} \overline \lambda'_{1}+
s'_{16} \lambda'_{2} \overline \lambda'_{1}$\cr
$c_{10}$&$s'_{13} \lambda'_{3} \overline \lambda'_{2}+
s'_{14} \lambda'_{4} \overline \lambda'_{2}-
s'_{15} \lambda'_{1} \overline \lambda'_{2}-
s'_{16} \lambda'_{2} \overline \lambda'_{2}$\cr
$c_{11}$&$s'_{17} \lambda'_{5} \overline \lambda'_{3}+
s'_{18} \lambda'_{5} \overline \lambda'_{6}$\cr
$c_{12}$&$s'_{19} \lambda'_{5} \overline \lambda'_{5}+
s'_{20} \lambda'_{5} \overline \lambda'_{4}$\cr
\omit&\omit \cr
\noalign{\hrule}
\omit&\omit \cr
$d_{1}$&$s'^3_{1}+s'^3_{3}+3s'_1s'^2_{2}+3s'_3s'^2_{4}$\cr
$d_{2}$&$s'^2_{1}s'_{13}+s'^2_{2}s'_{13}+
2s'_{1}s'_2s'_{14}+s'^2_{3}s'_{15}+s'^2_{4}s'_{15}+
2s'_{3}s'_4s'_{16}$\cr
$d_{3}$&$s'^3_{13}+s'^3_{15}+3s'_{13}s'^2_{14}+3s'_{15}s'^2_{16}$\cr
$d_{4}$&$s'^2_{13}s'_{1}+s'^2_{14}s'_{1}+
2s'_{13}s'_{14}s'_{2}+s'^2_{15}s'_{3}+s'^2_{16}s'_{3}+
2s'_{15}s'_4s'_{4}$\cr
$d_{5}$&$s'^3_{9}+s'^3_{11}+3s'_9s'^2_{10}+3s'_{11}s'^2_{12}$\cr
$d_{6}$&$s'^2_{11}s'_{9}+s'^2_{12}s'_{9}+
2s'_{10}s'_{11}s'_{12}+s'^2_{9}s'_{11}+s'^2_{10}s'_{11}+
2s'_{9}s'_{10}s'_{11}$\cr
$d_{7}$&$s'^2_{5}s'_{1}+s'^2_{6}s'_{1}+
2s'_{5}s'_6s'_{2}+s'^2_{7}s'_{3}+s'^2_{8}s'_{3}+
2s'_{7}s'_8s'_{4}$\cr
$d_{8}$&$s'^2_{5}s'_{13}+s'^2_{6}s'_{13}+
2s'_{5}s'_6s'_{14}+s'^2_{7}s'_{15}+s'^2_{8}s'_{15}+
2s'_{7}s'_8s'_{16}$\cr
$d_{9}$&$s'_{1}s'_{6}s'_{17}-s'_{2}s'_{5}s'_{17}+
s'_{3}s'_{7}s'_{18}-s'_{4}s'_{8}s'_{18}$\cr
$d_{10}$&$s'_{1}s'_{6}s'_{20}-s'_{2}s'_{5}s'_{20}+
s'_{3}s'_{7}s'_{19}-s'_{4}s'_{8}s'_{19}$\cr
$d_{11}$&$s'_{13}s'_{6}s'_{17}-s'_{14}s'_{5}s'_{17}+
s'_{15}s'_{7}s'_{18}-s'_{16}s'_{8}s'_{18}$\cr
$d_{12}$&$s'_{13}s'_{6}s'_{20}-s'_{14}s'_{5}s'_{20}+
s'_{15}s'_{7}s'_{19}-s'_{16}s'_{8}s'_{19}$\cr
\omit&\omit \cr
\noalign{\hrule}
\omit&\omit \cr
}}$$
\centerline{{\bf Table V}}

\vfill\eject

$$\vbox{
\tabskip=0.7truecm
\halign{#\hfil&#&#&#&#&#&#&#&#&#&#&#\cr
\noalign{\hrule}
\omit&\omit&\omit&\omit&\omit&\omit&
\omit&\omit&\omit&\omit&\omit&\omit\cr
&$l$&$e^c$&$h$&$h'$&$\nu^c$&$N$&$q$&$d'$&$u^c$&$d^c$&$d'^c$\cr
\omit&\omit&\omit&\omit&\omit&\omit&
\omit&\omit&\omit&\omit&\omit&\omit\cr
\noalign{\hrule}
\omit&\omit&\omit&\omit&\omit&\omit&
\omit&\omit&\omit&\omit&\omit&\omit\cr
$g_2$&$-1$&$-1$&$1$&$1$&$-1$&$1$&$-1$&$1$&$-1$&$-1$&$1$\cr
$g_3$&$1$&$\alpha^2$&$\alpha$&$\alpha^2$&$\alpha$&
$1$&$1$&$1$&$\alpha$&$\alpha^2$&$1$\cr
\omit&\omit&\omit&\omit&\omit&\omit&
\omit&\omit&\omit&\omit&\omit&\omit\cr
\noalign{\hrule}
\omit&\omit&\omit&\omit&\omit&\omit&
\omit&\omit&\omit&\omit&\omit&\omit\cr
}}$$
\centerline{{\bf Table VI}}

\vskip 1.truecm

$$\vbox{
\tabskip=0.7truecm
\halign{#\hfil&#&#&#&#&#&#&#\cr
\noalign{\hrule}
\omit&\omit&\omit&\omit&\omit&\omit&\omit&\omit\cr
&$l$&$e^c$&$h$&$h'$&$q$&$u^c$&$d^c$\cr
\omit&\omit&\omit&\omit&\omit&\omit&\omit&\omit\cr
\noalign{\hrule}
\omit&\omit&\omit&\omit&\omit&\omit&\omit&\omit\cr
$P_2$&$-1$&$-1$&$1$&$1$&$-1$&$-1$&$-1$\cr
$P_3$&$1$&$\alpha^2$&$\alpha^2$&$\alpha$&
$1$&$\alpha$&$\alpha^2$\cr
\omit&\omit&\omit&\omit&\omit&\omit&\omit&\omit\cr
\noalign{\hrule}
\omit&\omit&\omit&\omit&\omit&\omit&\omit&\omit\cr
}}$$
\centerline{{\bf Table VII}}

\vskip 1.truecm

$$
\vbox{
\tabskip=0truecm
\offinterlineskip
\halign{#&\quad\vrule#&\hfil\quad#\hfil&\quad\vrule#&
\hfil\quad#\quad\hfil&\vrule#&\quad\hfil #\hfil\quad&\vrule#\cr
&&$N$&&$\nu^c$&&$s$&\cr
\omit&height3pt&\omit&&\omit&&\omit&\cr
\omit&height3pt&\omit&&\omit&&\omit&\cr
\noalign{\hrule}
\omit&height3pt&\omit&&\omit&&\omit&\cr
\omit&height3pt&\omit&&\omit&&\omit&\cr
$P_3\;$&&$\lambda_{2}\;\;\lambda_{4}\;\;\lambda_{5}\;\;\lambda_{6}\;\;
\lambda_{7}$&
&$\lambda_{3}\;\;\lambda_{9}$&&$s_2\;\;s_4\;\;s_6\;\;
s_8\;\;s_{14}\;s_{16}$&\cr
\omit&height3pt&\omit&&\omit&&\omit&\cr
\omit&height3pt&\omit&&\omit&&\omit&\cr
&&$\overline \lambda_{1}\;\;\overline\lambda_{2}$&
&$\overline \lambda_{3}\;\;\overline \lambda_{4}$&&&\cr
\omit&height3pt&\omit&&\omit&&\omit&\cr
\omit&height3pt&\omit&&\omit&&\omit&\cr
\noalign{\hrule}
\omit&height3pt&\omit&&\omit&&\omit&\cr
\omit&height3pt&\omit&&\omit&&\omit&\cr
$P_2\;$&&$\;\lambda'_{1}\;\;\lambda'_{3}\;\;\lambda'_{5}\;\;
\lambda'_{6}\;\;\lambda'_{8}\;$&
&$\lambda'_{2}\;\;\lambda'_{4}\;\;\lambda'_{7}\;\;\lambda'_{9}$&
&$\; s'_1\;\;s'_3\;\;s'_5\;\;s'_7\;\;s'_{9}\;\;s'_{11}\;s'_{13}\;
s'_{15}\;$&\cr
\omit&height3pt&\omit&&\omit&&\omit&\cr
\omit&height3pt&\omit&&\omit&&\omit&\cr
&&$\overline \lambda'_{1}\;\;\overline\lambda'_{2}$&
&$\;\overline \lambda'_{3}\;\;\overline \lambda'_{4}\;\;
\overline \lambda'_{5}\;\;\overline \lambda'_{6}\;\;$&&&\cr
\omit&height3pt&\omit&&\omit&&\omit&\cr
\omit&height3pt&\omit&&\omit&&\omit&\cr
\noalign{\hrule}
}}$$
\centerline{{\bf Table VIII}}

\vfill\eject\end

***********************************************************

\nopagenumbers
\hsize=18cm
\baselineskip=1cm
$$
\vbox{
\tabskip=0.3truecm
\offinterlineskip
\halign{\hfil#\hfil&\hfil#\hfil&\hfil#\hfil&\hfil#\hfil&\hfil#\hfil&\hfil
#\hfil&\hfil\vrule#\hfil&\hfil#\hfil&\hfil#\hfil&
\hfil#\hfil&\hfil#\hfil&\hfil\vrule\hfil#\hfil&
\hfil#\hfil&\hfil#\hfil&\hfil#\hfil&\hfil#\hfil&\hfil#\hfil&
\hfil\vrule#\hfil&\hfil#\hfil\cr
&&$h'_2$&$h'_4$&$h'_7$&$h'_9$&\omit&$\overline h_3$&
$\overline h_4$&$\overline h_5$&$\overline h_6$&\omit&
$l_1$&$l_3$&$l_5$&$l_6$&$l_8$&\omit&$\lambda_g$\cr
\omit&\omit&\omit&\omit&\omit&\omit&height2pt&\omit&\omit&\omit&\omit&&
\omit&\omit&\omit&\omit&\omit&&\omit\cr
\omit&\omit&\omit&\omit&\omit&\omit&height2pt&\omit&\omit&\omit&\omit&&
\omit&\omit&\omit&\omit&\omit&&\omit\cr
$h_2$&&$x$&$$&$$&$$&&$$&$$&$$&$$&&$$&$y$&$$&$$&$$&&$$\cr
\omit&\omit&\omit&\omit&\omit&\omit&height2pt&\omit&\omit&\omit&\omit&&
\omit&\omit&\omit&\omit&\omit&&\omit\cr
$h_4$&&$$&$$&$x$&$x$&&$s$&$s$&$$&$$&&$y$&$$&$$&$$&$$&&$$\cr
\omit&\omit&\omit&\omit&\omit&\omit&height2pt&\omit&\omit&\omit&\omit&&
\omit&\omit&\omit&\omit&\omit&&\omit\cr
$h_7$&&$$&$x$&$$&$$&&$$&$$&$$&$$&&$$&$$&$y$&$$&$$&&\cr
\omit&\omit&\omit&\omit&\omit&\omit&height2pt&\omit&\omit&\omit&\omit&&
\omit&\omit&\omit&\omit&\omit&&\omit\cr
$h_9$&&$$&$x$&$$&$$&&$$&$$&$s$&$s$&&$$&$$&$$&$$&$$&&$$\cr
\omit&\omit&\omit&\omit&\omit&\omit&height2pt&\omit&\omit&\omit&\omit&&
\omit&\omit&\omit&\omit&\omit&&\omit\cr
\noalign{\hrule}
\omit&\omit&\omit&\omit&\omit&\omit&height2pt&\omit&\omit&\omit&\omit&&
\omit&\omit&\omit&\omit&\omit&&\omit\cr
$\overline h'_3$&&$$&$s$&$$&$$&&$$&$$&$x$&$$&&$$&$$&$$&$$&$$&&$y$\cr
\omit&\omit&\omit&\omit&\omit&\omit&height2pt&\omit&\omit&\omit&\omit&&
\omit&\omit&\omit&\omit&\omit&&\omit\cr
$\overline h'_4$&&$$&$s$&$$&$$&&$$&$$&$$&$x$&&$$&$$&$$&$$&$$&&$y$\cr
\omit&\omit&\omit&\omit&\omit&\omit&height2pt&\omit&\omit&\omit&\omit&&
\omit&\omit&\omit&\omit&\omit&&\omit\cr
$\overline h'_5$&&$$&$$&$$&$s$&&$x$&$$&$$&$$&&$$&$$&$$&$$&$$&&$$\cr
\omit&\omit&\omit&\omit&\omit&\omit&height2pt&\omit&\omit&\omit&\omit&&
\omit&\omit&\omit&\omit&\omit&&\omit\cr
$\overline h'_6$&&$$&$$&$$&$s$&&$$&$x$&$$&$$&&$$&$$&$$&$$&$$&&$$\cr
\omit&\omit&\omit&\omit&\omit&\omit&height2pt&\omit&\omit&\omit&\omit&&
\omit&\omit&\omit&\omit&\omit&&\omit\cr
\noalign{\hrule}
\omit&\omit&\omit&\omit&\omit&\omit&height2pt&\omit&\omit&\omit&\omit&&
\omit&\omit&\omit&\omit&\omit&&\omit\cr
$\overline l_1$&&$$&$$&$$&$$&&$$&$$&$y$&$y$&&$$&$r$&$$&$$&$$&&$x$\cr
\omit&\omit&\omit&\omit&\omit&\omit&height2pt&\omit&\omit&\omit&\omit&&
\omit&\omit&\omit&\omit&\omit&&\omit\cr
$\overline l_2$&&$$&$$&$$&$$&&$$&$$&$y$&$y$&&$$&$r$&$$&$$&$$&&\cr
\omit&\omit&\omit&\omit&\omit&\omit&height2pt&\omit&\omit&\omit&\omit&&
\omit&\omit&\omit&\omit&\omit&&\omit\cr
\noalign{\hrule}
\omit&\omit&\omit&\omit&\omit&\omit&height2pt&\omit&\omit&\omit&\omit&&
\omit&\omit&\omit&\omit&\omit&&\omit\cr
$\lambda'_g$&&$$&$$&$y$&$$&&$$&$$&$$&$$&&$x$&$$&$$&$$&$$&&\cr
}}$$
\centerline {\hfil(a)\hfil}

$$
\vbox{
\tabskip=0.3truecm
\offinterlineskip
\halign{\hfil#\hfil&\hfil#\hfil&\hfil#\hfil&\hfil#\hfil&
\hfil#\hfil&\hfil#\hfil&\hfil#\hfil&\hfil#\hfil&
\hfil#\hfil&\hfil#\hfil&\hfil#\hfil
&\hfil\vrule#\hfil&\hfil#\hfil&\hfil#\hfil\cr
&&${e^c}_1$&${e^c}_2$&${e^c}_3$&${e^c}_4$&${e^c}_5$&${e^c}_6$&
${e^c}_7$&${e^c}_8$&${e^c}_9$&\omit&$\lambda^1_g$&$\lambda^2_g$\cr
\omit&\omit&\omit&\omit&\omit&\omit&\omit&\omit&\omit&\omit&
\omit&height2pt&\omit&\omit\cr
\omit&\omit&\omit&\omit&\omit&\omit&\omit&\omit&\omit&\omit&
\omit&height2pt&\omit&\omit\cr
$\overline {e^c}_1$&&$$&$\cdot$&$r$&$\cdot$&$$&$$&$\cdot$&$$&
$\cdot$&&$x$&$\cdot$\cr
\omit&\omit&\omit&\omit&\omit&\omit&\omit&\omit&\omit&\omit&
\omit&height2pt&\omit&\omit\cr
$\overline {e^c}_2$&&$$&$\cdot$&$r$&$\cdot$&$$&$$&$\cdot$&$$&
$\cdot$&&$x$&$\cdot$\cr
\omit&\omit&\omit&\omit&\omit&\omit&\omit&\omit&\omit&\omit&
\omit&height2pt&\omit&\omit\cr
$\overline {e^c}_3$&&$\cdot$&$$&$\cdot$&$s$&$\cdot$&
$\cdot$&$$&$\cdot$&$$&&$\cdot$&$y$\cr
\omit&\omit&\omit&\omit&\omit&\omit&\omit&\omit&\omit&\omit&
\omit&height2pt&\omit&\omit\cr
$\overline {e^c}_4$&&$\cdot$&$$&$\cdot$&$s$&$\cdot$&
$\cdot$&$$&$\cdot$&$$&&$\cdot$&$y$\cr
\omit&\omit&\omit&\omit&\omit&\omit&\omit&\omit&\omit&\omit&
\omit&height2pt&\omit&\omit\cr
$\overline {e^c}_5$&&$\cdot$&$$&$\cdot$&$$&$\cdot$&
$\cdot$&$$&$\cdot$&$s$&&$\cdot$&$$\cr
\omit&\omit&\omit&\omit&\omit&\omit&\omit&\omit&\omit&\omit&
\omit&height2pt&\omit&\omit\cr
$\overline {e^c}_6$&&$\cdot$&$$&$\cdot$&$$&$\cdot$&
$\cdot$&$$&$\cdot$&$s$&&$\cdot$&$$\cr
\omit&\omit&\omit&\omit&\omit&\omit&\omit&\omit&\omit&\omit&
\omit&height2pt&\omit&\omit\cr
\noalign{\hrule}
\omit&\omit&\omit&\omit&\omit&\omit&\omit&\omit&\omit&\omit&
\omit&height2pt&\omit&\omit\cr
$\lambda'^1_g$&&$x$&$\cdot$&$$&$\cdot$&$$&$$&$\cdot$&$$&
$\cdot$&&$$&$\cdot$\cr
\omit&\omit&\omit&\omit&\omit&\omit&\omit&\omit&\omit&\omit&
\omit&height2pt&\omit&\omit\cr
$\lambda'^2_g$&&$\cdot$&$$&$\cdot$&$$&$\cdot$&
$\cdot$&$y$&$\cdot$&$$&&$\cdot$&$$\cr }}$$
\centerline {\hfil(b)\hfil}

$$
\vbox{
\tabskip=0.3truecm
\offinterlineskip
\halign{\hfil#\hfil&\hfil#\hfil&\hfil#\hfil&\hfil#\hfil&\hfil#\hfil&
\hfil#\hfil&\hfil#\hfil&\hfil\vrule#\hfil&\hfil#\hfil&\hfil#\hfil&
\hfil\vrule\hfil#\hfil&\hfil#\hfil&\hfil#\hfil&\hfil#\hfil&\hfil#\hfil\cr
&&$h'_1$&$h'_3$&$h'_5$&$h'_6$&$h'_8$&\omit&$\overline h_1$&
$\overline h_2$&\omit&$l_2$&$l_4$&$l_7$&$l_9$\cr
\omit&\omit&\omit&\omit&\omit&\omit&\omit&height2pt&
\omit&\omit&&\omit&\omit&\omit&\omit\cr
\omit&\omit&\omit&\omit&\omit&\omit&\omit&height2pt&
\omit&\omit&&\omit&\omit&\omit&\omit\cr
$h_1$&&$x$&$$&$$&$$&$$&&$$&$$&&$$&$y$&$$&$$\cr
\omit&\omit&\omit&\omit&\omit&\omit&\omit&height2pt&
\omit&\omit&&\omit&\omit&\omit&\omit\cr
$h_3$&&$$&$$&$x$&$$&$x$&&$r$&$r$&&$y$&$$&$$&$$\cr
\omit&\omit&\omit&\omit&\omit&\omit&\omit&height2pt&
\omit&\omit&&\omit&\omit&\omit&\omit\cr
$h_5$&&$$&$x$&$$&$$&$$&&$$&$$&&$$&$$&$y$&$$\cr
\omit&\omit&\omit&\omit&\omit&\omit&\omit&height2pt&
\omit&\omit&&\omit&\omit&\omit&\omit\cr
$h_6$&&$$&$$&$$&$$&$$&&$$&$$&&$$&$$&$$&$$\cr
\omit&\omit&\omit&\omit&\omit&\omit&\omit&height2pt&
\omit&\omit&&\omit&\omit\cr
$h_8$&&$$&$x$&$$&$$&$$&&$$&$$&&$$&$$&$$&$$\cr
\omit&\omit&\omit&\omit&\omit&\omit&\omit&height2pt&
\omit&\omit&&\omit&\omit&\omit&\omit\cr
\noalign{\hrule}
\omit&\omit&\omit&\omit&\omit&\omit&\omit&height2pt&
\omit&\omit&&\omit&\omit&\omit&\omit\cr
$\overline h'_1$&&$$&$r$&$$&$$&$$&&$x$&$$&&$$&$$&$$&$$\cr
\omit&\omit&\omit&\omit&\omit&\omit&\omit&height2pt&
\omit&\omit&&\omit&\omit&\omit&\omit\cr
$\overline h'_2$&&$$&$r$&$$&$$&$$&&$$&$x$&&$$&$$&$$&$$\cr
\omit&\omit&\omit&\omit&\omit&\omit&\omit&height2pt&
\omit&\omit&&\omit&\omit&\omit&\omit\cr
\noalign{\hrule}
\omit&\omit&\omit&\omit&\omit&\omit&\omit&height2pt&
\omit&\omit&&\omit&\omit&\omit&\omit\cr
$\overline l_3$&&$$&$$&$$&$$&$$&&$$&$$&&$$&$s$&$$&$$\cr
\omit&\omit&\omit&\omit&\omit&\omit&\omit&height2pt&
\omit&\omit&&\omit&\omit&\omit&\omit\cr
$\overline l_4$&&$$&$$&$$&$$&$$&&$$&$$&&$$&$s$&$$&$$\cr
\omit&\omit&\omit&\omit&\omit&\omit&\omit&height2pt&
\omit&\omit&&\omit&\omit&\omit&\omit\cr
$\overline l_5$&&$$&$$&$$&$$&$$&&$y$&$y$&&$$&$$&$$&$s$\cr
\omit&\omit&\omit&\omit&\omit&\omit&\omit&height2pt&
\omit&\omit&&\omit&\omit&\omit&\omit\cr
$\overline l_6$&&$$&$$&$$&$$&$$&&$y$&$y$&&$$&$$&$$&$s$\cr
}}$$
\centerline {\hfil(c)\hfil}
\centerline {\bf \hfil Fig. 1 \hfil}

\vfil\eject

$$
\vbox{
\tabskip=0.25truecm
\offinterlineskip
\halign{\hfil#\hfil&\hfil#\hfil&\hfil#\hfil&\hfil#\hfil&\hfil#\hfil&
\hfil#\hfil&\hfil#\hfil&\hfil#\hfil&\hfil#\hfil&\hfil\vrule#\hfil&
\hfil#\hfil&\hfil#\hfil&\hfil#\hfil&\hfil#\hfil&\hfil\vrule#\hfil&
\hfil#\hfil&\hfil#\hfil&\hfil#\hfil&\hfil#\hfil&\hfil#\hfil&\hfil#\hfil&
\hfil#\hfil\cr
&&$d'^c_1$&$d'^c_2$&$d'^c_3$&$d'^c_4$&$d'^c_5$&$d'^c_6$&$d'^c_7$&\omit&
$\overline{d'}_1$&$\overline{d'}_2$&$\overline{d'}_3$&$\overline{d'}_4$&
\omit&$d^c_1$&$d^c_2$&$d^c_3$&$d^c_4$&$d^c_5$&$d^c_6$&$d^c_7$\cr
\omit&\omit&\omit&\omit&\omit&\omit&\omit&\omit&\omit&height2pt&
\omit&\omit&\omit&\omit&height2pt&
\omit&\omit&\omit&\omit&\omit&\omit&\omit\cr
\omit&\omit&\omit&\omit&\omit&\omit&\omit&\omit&\omit&height2pt&
\omit&\omit&\omit&\omit&height2pt&
\omit&\omit&\omit&\omit&\omit&\omit&\omit\cr
$d'_1$&&$$&$$&$$&$$&$\cdot$&$$&$\cdot$&&$r$&$\cdot$&$r$&$\cdot$&&
$\cdot$&$\cdot$&$\cdot$&$\cdot$&$$&$\cdot$&$$\cr
\omit&\omit&\omit&\omit&\omit&\omit&\omit&\omit&\omit&height2pt&
\omit&\omit&\omit&\omit&height2pt&
\omit&\omit&\omit&\omit&\omit&\omit&\omit\cr
$d'_2$&&$$&$$&$$&$$&$\cdot$&$x$&$\cdot$&&$s$&$\cdot$&$s$&$\cdot$&&
$\cdot$&$\cdot$&$\cdot$&$\cdot$&$$&$\cdot$&$$\cr
\omit&\omit&\omit&\omit&\omit&\omit&\omit&\omit&\omit&height2pt&
\omit&\omit&\omit&\omit&height2pt&
\omit&\omit&\omit&\omit&\omit&\omit&\omit\cr
$d'_3$&&$$&$$&$$&$x$&$\cdot$&$$&$\cdot$&&$$&$\cdot$&$$&$\cdot$&&
$\cdot$&$\cdot$&$\cdot$&$\cdot$&$$&$\cdot$&$$\cr
\omit&\omit&\omit&\omit&\omit&\omit&\omit&\omit&\omit&height2pt&
\omit&\omit&\omit&\omit&height2pt&
\omit&\omit&\omit&\omit&\omit&\omit&\omit\cr
$d'_4$&&$$&$$&$x$&$$&$\cdot$&$$&$\cdot$&&$s$&$\cdot$&$s$&$\cdot$&&
$\cdot$&$\cdot$&$\cdot$&$\cdot$&$$&$\cdot$&$y$\cr
\omit&\omit&\omit&\omit&\omit&\omit&\omit&\omit&\omit&height2pt&
\omit&\omit&\omit&\omit&height2pt&
\omit&\omit&\omit&\omit&\omit&\omit&\omit\cr
$d'_5$&&$\cdot$&$\cdot$&$\cdot$&$\cdot$&$$&$\cdot$&$$&&
$\cdot$&$s$&$\cdot$&$s$&&$$&$$&$$&$$&$\cdot$&$y$&$\cdot$\cr
\omit&\omit&\omit&\omit&\omit&\omit&\omit&\omit&\omit&height2pt&
\omit&\omit&\omit&\omit&height2pt&
\omit&\omit&\omit&\omit&\omit&\omit&\omit\cr
$d'_6$&&$$&$x$&$$&$$&$\cdot$&$x$&$\cdot$&&$r$&$\cdot$&$r$&$\cdot$&&
$\cdot$&$\cdot$&$\cdot$&$\cdot$&$y$&$\cdot$&$$\cr
\omit&\omit&\omit&\omit&\omit&\omit&\omit&\omit&\omit&height2pt&
\omit&\omit&\omit&\omit&height2pt&
\omit&\omit&\omit&\omit&\omit&\omit&\omit\cr
$d'_7$&&$\cdot$&$\cdot$&$\cdot$&$\cdot$&$$&$\cdot$&$x$&&
$\cdot$&$r$&$\cdot$&$r$&&$$&$$&$$&$y$&$\cdot$&$$&$\cdot$\cr
\omit&\omit&\omit&\omit&\omit&\omit&\omit&\omit&\omit&height2pt&
\omit&\omit&\omit&\omit&height2pt&
\omit&\omit&\omit&\omit&\omit&\omit&\omit\cr
\noalign{\hrule}
\omit&\omit&\omit&\omit&\omit&\omit&\omit&\omit&\omit&height2pt&
\omit&\omit&\omit&\omit&height2pt&
\omit&\omit&\omit&\omit&\omit&\omit&\omit\cr
$\overline{d'^c}_1$&&$r$&$s$&$$&$s$&$\cdot$&$r$&
$\cdot$&&$x$&$\cdot$&$$&$\cdot$&&
$\cdot$&$\cdot$&$\cdot$&$\cdot$&$$&$\cdot$&$$\cr
\omit&\omit&\omit&\omit&\omit&\omit&\omit&\omit&\omit&height2pt&
\omit&\omit&\omit&\omit&height2pt&
\omit&\omit&\omit&\omit&\omit&\omit&\omit\cr
$\overline{d'^c}_2$&&$\cdot$&$\cdot$&$\cdot$&$\cdot$&$s$&$\cdot$&$r$&&
$\cdot$&$x$&$\cdot$&$$&&$$&$$&$$&$$&$\cdot$&$$&$\cdot$\cr
\omit&\omit&\omit&\omit&\omit&\omit&\omit&\omit&\omit&height2pt&
\omit&\omit&\omit&\omit&height2pt&
\omit&\omit&\omit&\omit&\omit&\omit&\omit\cr
$\overline{d'^c}_3$&&$r$&$s$&$$&$s$&$\cdot$&$r$&
$\cdot$&&$$&$\cdot$&$x$&$\cdot$&&
$\cdot$&$\cdot$&$\cdot$&$\cdot$&$$&$\cdot$&$$\cr
\omit&\omit&\omit&\omit&\omit&\omit&\omit&\omit&\omit&height2pt&
\omit&\omit&\omit&\omit&height2pt&
\omit&\omit&\omit&\omit&\omit&\omit&\omit\cr
$\overline{d'^c}_4$&&$\cdot$&$\cdot$&$\cdot$&$\cdot$&$s$&$\cdot$&$r$&&
$\cdot$&$$&$\cdot$&$x$&&$$&$$&$$&$$&$\cdot$&$$&$\cdot$\cr
\omit&\omit&\omit&\omit&\omit&\omit&\omit&\omit&\omit&height2pt&
\omit&\omit&\omit&\omit&height2pt&
\omit&\omit&\omit&\omit&\omit&\omit&\omit\cr
\noalign{\hrule}
\omit&\omit&\omit&\omit&\omit&\omit&\omit&\omit&\omit&height2pt&
\omit&\omit&\omit&\omit&height2pt&
\omit&\omit&\omit&\omit&\omit&\omit&\omit\cr
$\overline{d^c}_1$&&$\cdot$&$\cdot$&$\cdot$&$\cdot$&$$&$\cdot$&$$&&
$\cdot$&$$&$\cdot$&$$&&$r$&$s$&$$&$s$&$\cdot$&$r$&$\cdot$\cr
\omit&\omit&\omit&\omit&\omit&\omit&\omit&\omit&\omit&height2pt&
\omit&\omit&\omit&\omit&height2pt&
\omit&\omit&\omit&\omit&\omit&\omit&\omit\cr
$\overline{d^c}_2$&&$$&$$&$$&$$&$\cdot$&$$&$\cdot$&&$$&$\cdot$&$$&$\cdot$&&
$\cdot$&$\cdot$&$\cdot$&$\cdot$&$s$&$\cdot$&$r$\cr
\omit&\omit&\omit&\omit&\omit&\omit&\omit&\omit&\omit&height2pt&
\omit&\omit&\omit&\omit&height2pt&
\omit&\omit&\omit&\omit&\omit&\omit&\omit\cr
$\overline{d^c}_3$&&$\cdot$&$\cdot$&$\cdot$&$\cdot$&$$&$\cdot$&$$&&
$\cdot$&$$&$\cdot$&$$&&$r$&$s$&$$&$s$&$\cdot$&$r$&$\cdot$\cr
\omit&\omit&\omit&\omit&\omit&\omit&\omit&\omit&\omit&height2pt&
\omit&\omit&\omit&\omit&height2pt&
\omit&\omit&\omit&\omit&\omit&\omit&\omit\cr
$\overline{d^c}_4$&&$$&$$&$$&$$&$\cdot$&$$&$\cdot$&&$$&$\cdot$&$$&$\cdot$&&
$\cdot$&$\cdot$&$\cdot$&$\cdot$&$s$&$\cdot$&$r$\cr
}}$$
\centerline {\hfil(a)\hfil}
\vskip 0.5truecm

\baselineskip=0.4cm

$$
\vbox{
\tabskip=0.25truecm
\halign{\hfil#\hfil&\hfil#\hfil&\hfil#\hfil&\hfil#\hfil&\hfil#\hfil&
\hfil#\hfil&\hfil#\hfil&\hfil#\hfil&\hfil#\hfil\cr
&&$q_1$&$q_2$&$q_3$&$q_4$&$q_5$&$q_6$&$q_7$\cr
\omit&\omit&\omit&\omit&\omit&\omit&\omit&\omit\cr
$\overline{q}_1$&&$r$&$s$&$$&$s$&$\cdot$&$r$&$\cdot$\cr
$\overline{q}_2$&&$\cdot$&$\cdot$&$\cdot$&$\cdot$&$s$&$\cdot$&$r$\cr
$\overline{q}_3$&&$r$&$s$&$$&$s$&$\cdot$&$r$&$\cdot$\cr
$\overline{q}_4$&&$\cdot$&$\cdot$&$\cdot$&$\cdot$&$s$&$\cdot$&$r$\cr
}}$$
\baselineskip=1cm
\centerline {\hfil(b)\hfil}
\centerline {\bf \hfil Fig. 2 \hfil}
\vfil\eject\end